\documentclass[fullpage]{article}
\usepackage{amsthm}
\usepackage{amsmath}
\usepackage{color}
\usepackage{url} 
\usepackage{algorithm}
\usepackage{algorithmicx}
\usepackage[noend]{algpseudocode}
\usepackage{amssymb}
\usepackage{bm}
\usepackage[margin=1in]{geometry}
\usepackage{comment}
\newtheorem{theorem}{Theorem}[section]
\newtheorem{lemma}[theorem]{Lemma}

\providecommand{\keywords}[1]
{
  \small	
  \textbf{\textit{Keywords---}} #1
}

\newpage
\title{On Updating and Querying Submatrices}
\author{Jason Yang, Jun Wan}
\date{}

\begin{document}
\maketitle
\begin{abstract}
Given a $d$-dimensional matrix, an \textit{update} changes each element in a given submatrix from $x$ to $x\bigtriangledown v$, where $v$ is a given constant. A \textit{query} returns the $\bigtriangleup$ of all elements in a given submatrix. We study the cases where $\bigtriangledown$ and $\bigtriangleup$ are both commutative and associative binary operators. When $d = 1$ and the matrix is an array of length $N$, updates and queries can be performed in $O(\log N)$ worst-case time for many $(\bigtriangledown,\bigtriangleup)$ by using a segment tree with lazy propagation. However, when $d\ge 2$, similar techniques usually cannot be generalized. We show that if min-plus matrix multiplication cannot be computed in $O(N^{3-\varepsilon})$ time for any $\varepsilon>0$, then for $(\bigtriangledown,\bigtriangleup)=(+,\min)$, either updates or queries cannot both run in $O(N^{1-\varepsilon})$ time for any constant $\varepsilon>0$, or preprocessing cannot run in polynomial time. Finally, we show a special case where lazy propagation can be generalized for $d\ge 2$ and where updates and queries can run in $O(\log^d N)$ worst-case time. We present an algorithm that meets this running time and is slightly simpler than similar algorithms of previous works.
\end{abstract}


\tableofcontents
\newpage
\section{Introduction}
Range queries occur often in many computer science problems, such as computational geometry \cite{orth} and image retrieval \cite{img}. Many problems can be reduced to range queries: for example, finding the least common ancestors of two nodes in a tree can be reduced to performing a range minimum query on an array \cite{lca}.

Range queries can also be combined with range updates to create a dynamic range query problem. In this paper, we study a specific dynamic range query problem that we call the $d$-dimensional submatrix update-query problem: given two binary operators $\bigtriangledown$ and $\bigtriangleup$, the problem asks for an efficient data structure that supports two operations, update and query, on a $d$-dimensional hypermatrix. Both operations take a $d$-dimensional submatrix of the hypermatrix as input. The update operation combines each element in the submatrix with $\bigtriangledown$ and a constant $v$, which is also supplied by the update operation, and the query operation computes the result of combining all elements in the submatrix using $\bigtriangleup$. 

The segment tree, when combined with the  ``lazy value" technique, is a well-known solution to the case $d=1$, resulting in $O(\log N)$ time for both update and query operations over a $N$-length array, and it works over many different actions for updates (e.g. add, minimize, multiply, assign) and functions for queries (e.g. sum, minimum, product, k-th order statistic). The segment tree and similar data structures are also used in distributed hash tables \cite{dht} and various computational geometry problems, such as stabbing queries \cite{stab} and orthogonal range queries \cite{orth}, along with many other applications.

When $d=2$, however, the problem becomes much more difficult, as in general the lazy value technique cannot be generalized, and it is unknown whether sublinear-time updates and queries exist. In this paper, we provide impossibility results by reducing the two-dimensional update-query problem to min-plus matrix multiplication.
Since it is widely believed that no $O(N^{3-\varepsilon})$ time min-plus matrix multiplication algorithm exists for any constant $\varepsilon>0$, this strongly suggests that the update and query operations cannot both run in $O(N^{1-\varepsilon})$ time for any constant $\varepsilon>0$.

However, for special operator pairs $(\bigtriangledown,\bigtriangleup)$, like $(+,+)$ and $(*,*)$, there exist algorithms from \cite{fenw} and \cite{mult} that solve the 2D update-query problem with both updates and queries running in $O(\log N\log M)$ time. In this paper, we define the property these special operator pairs satisfy and give a simpler algorithm that achieves updates and queries under the same asymptotic running times as those from previous works.

In Section 2, we give an overview of the 1D segment tree. In Section 3, we reduce matrix multiplication to the 2D update-query problem. In Section 4, we present an efficient algorithm for special instances of the 2D update-query problem.

\subsection{Problem Definition}
In this subsection, we give a formal definition of the $d$-dimention update-query problem. 
Suppose $A$ is a $d$-dimensional hypermatrix. 
A $d$-dimensional submatrix $[l_0,r_0][l_1,r_1]\cdots[l_{d-1},r_{d-1}]$ covers all elements $A[x_0][x_1]\cdots[x_{d-1}]$ s.t. $l_i\leq x_i\leq r_i$ for all $0\leq i<d$ and contains all coordinates $(x_0,x_1,\cdots x_{d-1})$ where $l_i\leq x_i\leq r_i$. The update and query binary operators are $\bigtriangledown$ and $\bigtriangleup$ respectively.

The \textbf{\textit{update}} operation $U(B,v)$ modifies all elements in $B$ with $\bigtriangledown$ and $v$: 

\[U(B,v):=\forall (x_0,x_1,\cdots x_{d-1})\in B, \ A[x_0][x_1]\cdots[x_{d-1}]\gets A[x_0][x_1]\cdots[x_{d-1}]\bigtriangledown v\]

The \textbf{\textit{query}} operation $Q(B)$ returns the $\bigtriangleup$ of all elements in $B$:

\[Q(B):=\bigtriangleup_{(x_0,x_1,\cdots x_{d-1})\in B} A[x_0][x_1]\cdots[x_{d-1}]\]

Lastly, \textbf{\textit{preprocessing}} is the initialization of a 2D update-query data structure with an initial matrix $A$.

For this paper, we study the case where the operators $\bigtriangledown$ and $\bigtriangleup$ are over a set of objects $\mathcal{S}$, where $\mathcal{S}$ is either $\mathbb{R}\cup \{+\infty\}$, $\mathbb{R}\cup \{-\infty\}$, or $\mathbb{R}\cup \{-\infty,+\infty\}$, and where the operators satisfy three properties:

\begin{itemize}
    \item they are both commutative and associative
    \item $\exists$ identity values $\emptyset_U,\emptyset_Q \in\mathcal{S}$ s.t. $\forall a\in \mathcal{S}, a\bigtriangledown \emptyset_U=\emptyset_U \bigtriangledown a=a$ and $a\bigtriangleup \emptyset_Q=\emptyset_Q \bigtriangleup a=a$
    \item $\exists$ a function $F(a,v,k)$ s.t. for all sequences of numbers $[a_0,a_1,\cdots a_{k-1}]$, $\bigtriangleup_{0\leq i<k} (a_i \bigtriangledown v)\\=F(\bigtriangleup_{0\leq i<k} a_i,v,k)$, and $F(a,v,k)$ can be evaluated in constant time if its inputs are already known, ex.
    \begin{itemize}
        \item $(\bigtriangledown,\bigtriangleup)=(+,\min)\rightarrow F(a,v,k)=a+v$
        \item $(\bigtriangledown,\bigtriangleup)=(+,+)\rightarrow F(a,v,k)=a+v*k$
    \end{itemize}
\end{itemize}

\subsection{Preliminary Algebra}
\label{propertiesofF}
Here are two important properties of $F()$, which are crucial in the correctness of the algorithms we present in this paper. As a reminder, $F()$ is the function s.t. for all sequences of numbers $[a_0,a_1,\cdots a_{k-1}]$, $\bigtriangleup_{0\leq i<k} (a_i \bigtriangledown v)=F(\bigtriangleup_{0\leq i<k} a_i,v,k)$.

\begin{enumerate}
    \item If we have a sequence of numbers, $\bigtriangledown$ all of them with a value $x$, and then $\bigtriangledown$ all of the resulting numbers with a value $y$, it is the same as if we switched $x$ and $y$; it is also the same as if we $\bigtriangledown$ all of the numbers in the original sequence with $x\bigtriangledown y$. This is because $\bigtriangledown$ is commutative and associative.
    \item If we have two sequences of numbers, $\bigtriangledown$ all numbers in both sequences with the same value $v$, $\bigtriangleup$ all the numbers in each sequences, and $\bigtriangleup$ the resulting two numbers together, it will be the same as if we treated both sequences as one big sequence, $\bigtriangledown$ all numbers with $v$, and $\bigtriangleup$ all numbers in that big sequence.
\end{enumerate}

Formally, for sequences $S_0=[a_0,a_1,\cdots a_{p-1}], S_1=[b_0,b_1,\cdots b_{q-1}]$ and numbers $V_0=\bigtriangleup_{0\leq i<p} a_i,V_1=\bigtriangleup_{0\leq i<q} b_i$, these two properties can be written as

\begin{equation}
\begin{aligned}
\bigtriangleup_{0\leq i<p} (a_i\bigtriangledown x\bigtriangledown y)=\bigtriangleup_{0\leq i<p}(a_i\bigtriangledown (x\bigtriangledown y))&=F(V_0,x\bigtriangledown y,p) \\
=\bigtriangleup_{0\leq i<p}((a_i\bigtriangledown x)\bigtriangledown y)&=F(F(V_0,x,p),y,p) \\
=\bigtriangleup_{0\leq i<p}((a_i\bigtriangledown y)\bigtriangledown x)&=F(F(V_0,y,p),x,p)
\end{aligned}
\end{equation}

and

\begin{equation}
\begin{aligned}
(\bigtriangleup_{0\leq i<p} (a_i\bigtriangledown v))\bigtriangleup (\bigtriangleup_{0\leq i<q} (b_i\bigtriangledown v))&=F(V_0,v,p)\bigtriangleup F(V_1,v,q) \\
=\bigtriangleup_{0\leq i<p+q} ((S_0|S_1)[i]\bigtriangledown v)&=F(V_0\bigtriangleup V_1,v,p+q),
\end{aligned}
\end{equation}

where $(S_0|S_1)=[a_0,a_1,\cdots a_{p-1},b_0,b_1,\cdots b_{q-1}]$.

\subsection{Notation}
Here is some notation shorthand we use throughout this paper:
\begin{itemize}
    \item A 1D coordinate $(x)$ may be written as simply a number $x$
    \item $A[x_0][x_1]\cdots[x_{d-1}]$ may be written as $A(c)$, where $c=(x_0,x_1,\cdots x_{d-1})$
    \item For two 1D submatrices (subarrays) $a=[x_0,x_1],b=[y_0,y_1]$, the 2D submatrix $B=[x_0,x_1][y_0,y_1]$ may be expressed as $a*b$
    \item For a submatrix $B=[x_0,x_1][y_0,y_1]$, we define the attributes $B_X:=[x_0,x_1]$ and $B_Y:=[y_0,y_1]$
    \item A \textbf{node} is an object that stores information about a specific subset of the elements of a hypermatrix. We say that a node \textbf{covers} these elements, and also covers the coordinates of those elements.
    \item If there is a set operation written with or between nodes, those nodes are treated as the sets of elements (or coordinates) they cover, e.g.
    \begin{itemize}
        \item $|n|$ is the number of elements covered by $n$
        \item $m\cap n$ is the intersection of the sets of elements covered by $m$ and $n$
        \item $x\in n$ means that the coordinate $x$ is covered by $n$
        \item $n_X$ is the $_X$ attribute of the submatrix that $n$ covers (only defined if $n$ covers a submatrix)
    \end{itemize}
    \item $m\cap\not\subseteq n$ is short for $(m\cap n\neq \emptyset)\wedge (m\not\subseteq n)$
\end{itemize}

\section{The 1D Problem and the 1D Segment Tree}
In this section, we give an overview of the 1D segment tree, a well-known efficient solution to the 1D update-query problem that can perform any $\bigtriangledown$ update and any $\bigtriangleup$ query in $O(\log N)$ time over a $N$-length array. (The 1D segment tree even works for some non-commutative and non-associative $\bigtriangledown$ and $\bigtriangleup$, although the update and query algorithms for those operators are more complicated). We describe the intuition in Section 2.1, show the ``lazy value" technique in Section 2.2, and present the detailed update and query algorithms in Section 2.3.

A segment tree over a $N$-length array can be defined as follows:

\begin{enumerate}
    \item There is a ``root" node, which covers $[0,N-1]$ (i.e. the entire array).
    \item Every node $n$ that covers $[l,r]$, where $l<r$, contains two children nodes, $n_l$ and $n_r$, which cover $[l,m]$ and $[m+1,r]$ respectively, where $m=\lfloor(l+r)/2\rfloor$. We say that $n$ is the parent of $n_l$ and $n_r$.
    \item Every node $n$ has a value $n_V$ s.t. $n_V=\bigtriangleup_{x\in n} A[x]$ at initialization.
\end{enumerate}

We can associate every node with a ``depth": the root node has a depth of 0, and for any node with a depth of $d$, its children have a depth of $d+1$. It can be shown that the maximum depth of any node, i.e. the depth of the entire tree, is $O(\log N)$.
Moreover, there are exactly $2N-1$ total nodes (even when $N$ is not a power of 2), and that the values of all nodes can be initialized in $O(N)$ time in a bottom-up fashion: for each node $n$ from largest to smallest depth, set $n_V$ to $A[i]$ if $n$ covers $[i,i]$; otherwise, set $n_V$ to $(n_l)_V\bigtriangleup (n_r)_V$.

\subsection{Fast queries but slow updates}
A useful property of the segment tree is that for any subarray $R$, there exists a set of nodes $S(R)=\{n_0, n_1, \ldots n_{k-1}\}$ s.t. (1) $|S(R)|=O(\log N)$, (2) $\cup_{n\in S(R)}n=R$, and (3) all $n_i$ are mutually disjoint \cite{cp-seg}.
This property immediately yields a $O(\log N)$-time query algorithm, since $Q(R)=\bigtriangleup_{x\in R} A[x]=\bigtriangleup_{n\in S(R)} n_V$.

To perform the update operation $U(R,v)$, a typical approach is to change the node value $n_V$ for all nodes that intersect with $R$; however, there can be as many as $\Omega(N)$ such nodes in the worst case, leading to a $\Omega(N)$-time update algorithm. One way of solving this problem is to use the ``lazy value" technique, which we describe in the next section.

\subsection{The lazy value technique}
Instead of performing the update operation on any arbitrary subarray, we first study how to update the subarray covered by some node $n$; we call this \textit{updating the region of $n$}. This makes analysis easier because all nodes that intersect $n$ either subset it or superset it, since the nodes of the segment tree obey a \textit{tree structure} (which we define in Appendix \ref{quadtree}).

Note that while only $O(\log N)$ nodes can superset $n$, up to $\Omega(N)$ nodes can subset $n$ in the worst case, so these nodes are the most problematic. When we update the region of $n$ with the value $v$, $A[x]$ becomes $A[x]\bigtriangledown v$ for all $x\in n$, so for each node $m\subseteq n$, $m_V$ will change from $\bigtriangleup_{x\in m}A[x]$ to $\bigtriangleup_{x\in m}(A[x]\bigtriangledown v)$; by definition of $F()$, this means that $m_V$ will change to $F(m_V,v,|m|)$.

Because doing this action can take a lot of time, we instead defer it to a ``lazy" value $n_Z$, which represents the command ``$\forall m\subseteq n$, change $m_V$ to $F(m_V,n_Z,|m|)$". At initialization, all $n_Z=\emptyset_U$; when updating the region of $n$, we take care of all nodes $\subseteq n$ by changing $n_Z$ to $n_Z\bigtriangledown v$; this also allows us to stack multiple updates on the region of $n$, since $\bigtriangledown$ is associative. We say that the lazy value $n_Z$ \textit{covers} all $m\subseteq n$, including $n$.

After this, we still have to change $m_V$ for all $m$ that intersect with $n$ but are not covered by $n_Z$. One approach might be to split the elements that were combined together with $\bigtriangleup$ into the group covered by $n$ and the group not covered by $n$ and reconstruct $m_V$ that way, but there is a better approach, which comes from examining the query operation.

When querying the region of $n$, we cannot simply return $n_V$ anymore, as we need to also account for all $m_Z$ that cover $n$; these come from the nodes $m\supseteq n$. Since $m_Z$ contains the command ``change $n_V$ to $F(n_V,m_Z,|n|)$", stacking these commands together and using the properties of $F()$ means that the answer to the query is $F(n_V,\bigtriangledown_{m\supseteq n} m_Z,|n|)$; we call this the ``true" value of $n$, or $T(n)$.

To see how this helps us complete the update algorithm, note that every node $n$ that has children must satisfy $T(n)=T(n_l)\bigtriangleup T(n_r)$, because by definition, the region $n$ covers is the union of the regions that $n_l$ and $n_r$ cover. Substituting in the definition of $T()$,

\[F\Big(n_V,\bigtriangledown_{m\supseteq n} m_Z,|n|\Big)
= F\Big((n_l)_V,\bigtriangledown_{m\supseteq n_l} m_Z,|n_l|\Big)\bigtriangleup 
F\Big((n_r)_V,\bigtriangledown_{m\supseteq n_r} m_Z,|n_r|\Big).
\]

Letting $L=\bigtriangledown_{m\supseteq n} m_Z$ and using the properties of $F()$ shown in Section \ref{propertiesofF}, we deduce that

\begin{equation}
\begin{aligned}
F\Big(n_V,L,|n|\Big)& 
= F\Big((n_l)_V,(n_l)_Z\bigtriangledown L,|n_l|\Big)\bigtriangleup F\Big((n_r)_V,(n_r)_Z\bigtriangledown L,|n_r|\Big) \\
& = F\Big(F((n_l)_V,(n_l)_Z,|n_l|),L,|n_l|\Big)\bigtriangleup F\Big(F((n_r)_V,(n_r)_Z,|n_r|),L,|n_r|\Big) \\
& = F\Big(F((n_l)_V,(n_l)_Z,|n_l|)\bigtriangleup F((n_r)_V,(n_r)_Z,|n_r|),L,|n|\Big) \\
\Leftrightarrow n_V&=F\Big((n_l)_V,(n_l)_Z,|n_l|\Big)\bigtriangleup F\Big((n_r)_V,(n_r)_Z,|n_r|\Big)
\end{aligned}
\end{equation}

Thus, after changing $n_Z$, we simply restore this rule to $m_V$ for all $m\cap\not\subseteq n$ by setting $m_V$ to $F((m_l)_V,\allowbreak(m_l)_Z,|m_l|)\bigtriangleup F((m_r)_V,(m_r)_Z,|m_r|)$. To make sure that $(m_l)_V,(m_l)_Z,(m_r)_V$, and $(m_r)_Z$ are all up-to-date before using them to update $m_V$, we first update $m_V$ with $m$ as the parent of $n$, then with $m$ as the parent's parent of $n$, etc. in that specific order. Since there are only $O(\log N)$ such $m$, updating the region of a node takes $O(\log N)$ time.

\subsection{Fast queries and fast updates with the lazy value technique}
Since we can update and query the region of a single node in $O(\log N)$ time, we can update and query an arbitrary subarray in $O(\log^2 N)$ time. This is not optimal, as the nodes $\in S(R)$ for a subarray $R$ share many common ancestors, so these ancestors will be visited multiple times in the process. The following recursive algorithms of $U()$ and $Q()$ (which are more similar to traditional segment tree implementations), shown below, improve updates and queries to $O(\log N)$ time by visiting nodes in a specific order and implicitly using a property of $F()$.
\begin{algorithm}[H]
\begin{algorithmic}
\caption{Update/query a subarray}
\Function{U}{$R,v$}
    \State UN$($root node$,R,v)$
\EndFunction
\Function{UN}{$n,R,v$}
    \If {$n\subseteq R$}
        \State $n_Z\leftarrow n_Z\bigtriangledown v$
    \ElsIf {$n$ intersects with $R$}
        \State UN$(n_l,R,v)$
        \State UN$(n_r,R,v)$
        \State $n_V\leftarrow F((n_l)_V,(n_l)_Z,|n_l|)\bigtriangleup F((n_r)_V,(n_r)_Z,|n_r|)$
    \EndIf
\EndFunction
\Function{Q}{$R$}
    \State \Return QN$($root node$,R)$
\EndFunction
\Function{QN}{$n,R$}
    \If {$n\subseteq R$}
        \State \Return $F(n_V,n_Z,|n|)$
    \ElsIf {$n$ intersects with $R$}
        \State \Return $F($QN$(n_l,R)\bigtriangleup$QN$(n_r,R),n_Z,|n\cap R|)$
    \Else
        \State \Return $\emptyset_Q$
    \EndIf
\EndFunction
\end{algorithmic}
\end{algorithm}

The $UN(n,R,v)$ procedure can be thought of as ``update $n$ and all necessary nodes that subset $n$", and the $QN(n,R)$ procedure can be thought of as ``return the $\bigtriangleup$ of all elements covered by both $n$ and $R$, ignoring the lazy values $m_Z$ for all $m\supset n$". If $R$ is restricted to be exactly the region of a node, then the $UN()$ and $QN()$ algorithms simplify into the procedures we showed in the previous subsection for updating the region of a single node. We prove the correctness of these algorithms in Appendix \ref{correctsegtree}.

\section{Impossibility results for the general 2D problem}
Since the segment tree achieves $O(\log N)$-time updates and queries over a $N$-length array, it seems promising to generalize it to 2D, where it can hopefully achieve $O(\log^2 N)$-time updates and queries over a $N\times N$ matrix. Unfortunately, this is most likely not the case in general, as in this section we show that many variants of $N\times N$ matrix multiplication, such as standard and min-plus, can be solved with a 2D update-query data structure, using $O(N^2)$ many updates and queries. Thus, if it is possible to have a 2D update-query data structure perform updates and queries in $O(poly(\log(N)))$ time, then it will be possible to perform many kinds of matrix multiplication in $\tilde{O}(N^2)$ time; this seems highly unlikely to be possible, since there is currently no known min-plus multiplication algorithm that runs in faster than $O(N^{3-\varepsilon})$ time.

In Section 3.1, we show that the lazy value technique is difficult to generalize to 2 dimensions. In Section 3.2, we present the reduction to matrix multiplication, suggesting that updates and queries cannot both run in truly sublinear time, even if polynomial-time preprocessing is allowed.

\subsection{Intuition of Hardness}
In this subsection, we give some intuition on why the 2D problem is difficult, by showing that two possible generalizations of the 1D segment tree and the lazy value technique fail to perform both updates and queries in $O(poly(\log(N)))$ time. We also prove that other generalizations, such as quadtrees and arbitrary sets of nodes, also fail to meet this running time, in Appendix \ref{nodesbad}.

\subsubsection{Attempt 1}
One generalization of the 1D segment tree is to use a ``segment tree of segment trees": make a 1D segment tree over the rows of a matrix, and then for each node in that tree, make another 1D segment tree that goes over the columns.

Formally, let $A$ be a $N\times M$ matrix and $T$ be a 1D segment tree over a $N$-length array. Each node $n\in T$ has a 1D segment tree $n_A$ s.t. $n_A.Q(R)=\bigtriangleup_{x\in n, y\in R} A[x][y]$ for all subarrays $R$. Then $Q(B)=\bigtriangleup_{n\in T.S(B_X)}n_A.Q(B_Y)$, which runs in $O(\log N\log M)$ time.

However, updates are tricky to perform efficiently. For simplicity, suppose we update the submatrix $B$ with $B_X=n$ for some node $n$, with the value $v$. Then we would have to do $m_A.U(B_Y,v)$ for all $m\subseteq n$ (this is not exactly a $\bigtriangledown$ update; more on that in Appendix \ref{general2dsegtree}), which would already force the update operation to visit $\Omega(N)$ nodes in the worst case.

We could try to defer this action to a lazy value, $n_Z$, which stores the subarray $B_Y$, but lazy values now become hard to stack. This happens because unlike in 1D, where each lazy value only stores a single number, here we must store subarrays, and as more and more subarrays stack up, the aggregate command that all of them represent can become arbitrarily complex.

Dealing with the nodes $m\cap\not\subseteq n$ is also problematic: we cannot quickly calculate $(m_l)_A\bigtriangleup (m_r)_A$, as the two objects are segment trees, not numbers, and doing so will take $O(M)$ time. Additionally, we cannot apply a uniform subarray update over $m_A$. To see why, consider doing the update $U([0,0][i,j])$ on a 2x$M$ matrix $A$; the sequence $[A[0][i]\bigtriangleup A[1][i], \ 0\leq i<M]$ can change in arbitrarily complex ways, as any subset of elements in this sequence can change or stay the same. Most operator pairs $(\bigtriangledown,\bigtriangleup)$ lead to this inconvenient property, such as $(+,\min)$ and $(*,+)$.

In Appendix \ref{general2dsegtree}, we show that this idea of a ``segment tree of segment trees" leads to an algorithm that can perform updates in $O(N\log M+M\log N)$ time and queries in $O(\log N\log M)$ time.

\subsubsection{Attempt 2}
A different way of making a 2D segment tree over a $N\times M$ matrix $A$ involves looking at what the innermost nodes of the ``segment tree of segment trees" cover. Here each node covers a submatrix:

\begin{enumerate}
    \item For each 1D node $a\in T_X,b\in T_Y$, where $T_X$ and $T_Y$ are 1D segment trees over $N$-length and $M$-length arrays respectively, there is a 2D node $n$ that covers $a*b$
    \item Each 2D node $n$ has a value $n_V$ s.t. $n_V=\bigtriangleup_{(x,y)\in n} A[x][y]$ at initialization
    \item Each 2D node $n$ has a lazy value, $n_Z$, which represents the command ``for all 2D nodes $m\subseteq n$, change $m_V$ to $F(m_V,n_Z,|m|)$"
\end{enumerate}

In this model, a 2D segment tree is treated more like a ``product" of two 1D segment trees, rather than a 1D segment tree of 1D segment trees. As before, there are at most $O(NM)$ nodes and the 2D segment tree can be initialized in $O(NM)$ time in a bottom-up fashion.

Like before, queries can be performed quickly: any submatrix $B$ can be represented with the set of nodes $S(B)=\{n|n=p*q,\forall p\in T_X.S(B_X), q\in T_Y.S(B_Y)\}$; the ``true value" of any node $n$ is $F(n_V,\bigtriangledown_{m\supseteq n} m_Z,|n|)$, so $Q(B)=\bigtriangleup_{n\in S(B)}F(n_V,\bigtriangledown_{m\supseteq n} m_Z,|n|)$; since any node $n$ has at most $O(\log N \log M)$ nodes $m\supseteq n$ and $|S(B)|=|T_X.S(B_X)|*|T_Y.S(B_Y)|= O(\log N \log M)$, queries can be done in $O(\log^2 N \log^2 M)$ time.

Unfortunately, updates are once again trickier. For each node $n\in S(B)$, we must update $n_Z$ to $n_Z\bigtriangledown v$ and update $m_V$ for all $m\cap\not\subseteq n$; these consist of not only nodes $m\supset n$, but also nodes $m$ s.t. $(m_X\subseteq n_X)\wedge(m_Y\supset n_Y)$ or $(m_X\supset n_X)\wedge(m_Y\subseteq n_Y)$, which we call \textit{half-ancestors} of $n$. A node can have as many as $O(M\log N + N\log M)$ half-ancestors, so unfortunately this yields a $\tilde{O}(N+M)$ update algorithm.

\subsection{Reduction from matrix multiplication}

The two attempts in Section 3.1 imply that the 1D segment tree and lazy value technique are hard to generalize to the 2D problem. In this section, we reduce matrix multiplication to a series of 2D updates and queries, suggesting that no matter what algorithm is used, it is impossible to perform updates and queries in truly sublinear time for any polynomial-time preprocessing.

Define $(\bigtriangleup,\bigtriangledown)$ matrix multiplication of two $N\times N$ matrices $A$ and $B$, abbreviated as $A*B$, to be the matrix $C$ s.t. $C_{i,j}=\bigtriangleup_{0\leq k<N} (A_{i,k}\bigtriangledown B_{k,j})$ (notice that the order of the operations in the name is flipped in order to follow typical naming conventions for variants of matrix multiplication).

In this section, we prove that any $(\bigtriangleup,\bigtriangledown)$ matrix multiplication, with the only constraint being that there exists a function $inv()$ s.t. $x\bigtriangledown inv(x)=\emptyset_U$ for all $x$, can be computed using a 2D update-query data structure. Below is an algorithm that does this.

(Note: For multiplication, $inv(0)$ is undefined, but we can avoid this problem by appending an extra number to every element in the update-query data structure, which tracks the number of times that element has been multiplied by 0; this does not affect the asymptotic running time of updates and queries.)

\begin{algorithm}[H]
\caption{Perform the $(\bigtriangleup,\bigtriangledown)$ matrix product of $A$ and $B$, using a 2D update-query data structure}
\begin{algorithmic}
\State Let $A,B$ be any $N\times N$ matrices
\State Let $D$ be a 2D $(\bigtriangledown,\bigtriangleup)$ update-query data structure
\State Initialize $D$ with the matrix $A$
\State Let $C$ be a $N\times N$ matrix
\ForAll{$j\in [0,N-1]$}
    \ForAll{$i\in [0,N-1]$}
        \State $D.U([0,N-1][i,i],B_{i,j})$
    \EndFor
    \State (After these updates, $A_{x,y}$ has become $A_{x,y}\bigtriangledown B_{y,j}$)
    \ForAll{$i\in [0,N-1]$}
        \State $C_{i,j}\gets D.Q([i,i][0,N-1])$
    \EndFor
    \ForAll{$i\in [0,N-1]$}
        \State $D.U([0,N-1][i,i],inv(B_{i,j}))$
    \EndFor
\EndFor
\State \Return $C$
\end{algorithmic}
\end{algorithm}

Because there are $2N^2$ updates and $N^2$ queries, $(\bigtriangleup,\bigtriangledown)$ matrix multiplication can be done in $O(P(N)+N^2(T_U(N)+T_Q(N)))$ time, where $P(N)$, $T_U(N)$, and $T_Q(N)$ are the preprocessing, update, and query times respectively of the 2D update-query data structure over a $N\times N$ matrix.

If updates and queries can be done in polylogarithmic time, this would yield a $\tilde{O}(N^2)$-time algorithm. For comparison, the current best known algorithm for standard matrix multiplication runs in $O(N^{2.3728639})$ time \cite{fastmatmul}; for min-plus matrix multiplication, it is $O(\frac{N^3}{2^{\Omega(\log N)^{1/2}}})$ time \cite{minplus}. For the min-plus case, it seems unlikely that a general truly subcubic algorithm (i.e. a $O(N^{3-\varepsilon})$ algorithm where $\varepsilon>0$) exists; such algorithms \cite{specialminplus} only exist for restricted matrices; this will be important in the next subsection.

\subsubsection{Performing multiple matrix multiplications}
The current reduction proof does not rule out the possibility of fast updates and queries at the expense of slow (but still polynomial-time) preprocessing. However, we can expand Algorithm 2 to find matrix products of $K$ pairs of arbitrary matrices $(A_i,B_i)$, for all $0\leq i<K$, for any positive integer $K$, while only initializing the update-query data structure once:

\begin{algorithm}[H]
\caption{Perform $K$ $(\bigtriangleup,\bigtriangledown)$ matrix products of $A_i$ and $B_i$, using a 2D update-query data structure}
\begin{algorithmic}
\State Let $D$ be a 2D $(\bigtriangledown,\bigtriangleup)$ update-query data structure
\State Initialize $D$ with an arbitrary $N\times N$ matrix
\State Let each $C_k$, $0\leq k<K$ be a $N\times N$ matrix
\ForAll{$k\in [0,K-1]$}
    \ForAll{$i\in [0,N-1]$}\ForAll{$j\in [0,N-1]$}
            \State $D.U([i,i][j,j],inv(D.Q([i,i][j,j]))\bigtriangledown (A_k)_{i,j})$
    \EndFor\EndFor
    \ForAll{$j\in [0,N-1]$}
        \ForAll{$i\in [0,N-1]$}
            \State $D.U([0,N-1][i,i],(B_k)_{i,j})$
        \EndFor
        \ForAll{$i\in [0,N-1]$}
            \State $(C_k)_{i,j}\gets D.Q([i,i][0,N-1])$
        \EndFor
        \ForAll{$i\in [0,N-1]$}
            \State $D.U([0,N-1][i,i],inv((B_k)_{i,j}))$
        \EndFor
    \EndFor
\EndFor
\State \Return $C$
\end{algorithmic}
\end{algorithm}

From this algorithm, finding all $A_i*B_i$ takes $O(P(N)+KN^2(T_U(N)+T_Q(N)))$ time.

Although we believe that the optimal running time for performing $K$ multiplications of $N\times N$ matrices is $O(K*M(N))$, where $M(N)$ is the optimal time for multiplying two $N\times N$ matrices, we are unable to prove so. Instead, we note that the action of performing $K$ multiplications of $N\times N$ matrices can be used to perform one multiplication of two $K^{\frac{1}{3}}N\times K^{\frac{1}{3}}N$ matrices: write each of these large matrices as a $K^{\frac{1}{3}}\times K^{\frac{1}{3}}$ matrix of $N\times N$ matrices, perform a (matrix-times)-(matrix-plus) multiplication using the schoolbook algorithm (treating each $N\times N$ element matrix as if it was a number), and then ``unwrap" each $N\times N$ element matrix in the resulting matrix of matrices. Therefore, performing $K$ multiplications of $N\times N$ matrices must take at least $O(M(K^{\frac{1}{3}}N))$ time. Letting $m=K^\frac{1}{3}$, we have

\[O(P(N)+m^3N^2(T_U(N)+T_Q(N)))\geq O(M(mN))\]
\[\Rightarrow O(P(N))\geq O(M(mN)) \texttt{ or } O(T_U(N)+T_Q(N))\geq O(\frac{M(mN)}{m^3N^2})\]

Suppose that $M(N)>O(N^{3-\varepsilon})$ for all $\varepsilon>0$; then

\[O(P(N))> O(m^{3-\varepsilon}N^{3-\varepsilon}) \texttt{ or } O(T_U(N)+T_Q(N))> O(m^{-\varepsilon}N^{1-\varepsilon}).\]

If $m=N^c$ for some number $0<c<1$, then 

\[O(P(N))> O(N^{(3-\varepsilon)(1+c)}) \texttt{ or } O(T_U(N)+T_Q(N))> O(N^{1-\varepsilon-c\varepsilon}).\]

To prevent $O(N^{1-\varepsilon-c\varepsilon})$ from becoming $O(1)$, we require that $1-\varepsilon-c\varepsilon> 0 \Rightarrow c< \frac{1-\varepsilon}{\varepsilon}$. If $c=d\frac{1-\varepsilon}{\varepsilon}$ for some $d<1$, then

\[O(P(N))> O(N^{(3-\varepsilon)(1+d\frac{1-\varepsilon}{\varepsilon})}) \texttt{ or } O(T_U(N)+T_Q(N))> O(N^{(1-\varepsilon)(1-d)}).\]

Since $\lim_{\varepsilon\to 0^+} (3-\varepsilon)(1+d\frac{1-\varepsilon}{\varepsilon})=\infty$ for any $0<d<1$, this means that as $\varepsilon$ approaches 0, $O(T_U(N)+T_Q(N))\geq O(N^{1-d})$ or $O(P(N))$ is superpolynomial.

Therefore, if truly subcubic min-plus matrix multiplication is not possible, then at least one of the following is true: updates and queries cannot both run in truly sublinear time, or preprocessing time is superpolynomial. Given that min-plus matrix multiplication has been studied for several decades and no one has currently found a truly subcubic algorithm for general matrices \cite{minplus}, it seems reasonable to assume that no such algorithm exists, and to therefore assume our impossibility result holds.

This lower bound is essentially tight, since there exists a data structure called a \textit{quadtree}, which performs updates and queries in $O(N)$ time over a $N\times N$ matrix with polynomial-time preprocessing. We prove this in Appendix \ref{tightness}.

\section{$O(\log^2 N)$-time updates/queries for special operator pairs}
Given the hardness of the general 2D update-query problem presented in the previous section, it seems discouraging to try finding an efficient algorithm. However, if the operators $(\bigtriangledown,\bigtriangleup)$ satisfy a special property, this is not the case. In this section, we present an algorithm, partially inspired by \cite{animesh}, that can perform both updates and queries in $O(\log N\log M)$ time. The main idea is to use the 1D segment tree as a building block to make an efficient algorithm for a special case of the 2D update-query problem.

In Section 4.1, we define what we mean by ``special". In section 4.2, we present the detailed algorithms. In section 4.3, we show how the algorithms can be generalized to higher-dimensional variants of the update-query problem. Finally, in section 4.4, we compare our 2D update-query data structure to those of \cite{fenw} and \cite{mult}.

\subsection{Definition of ``special"}
By looking at the 2D segment tree, one may wonder if it is possible to deal with half-ancestors with additional kinds of lazy values. For most operator pairs $(\bigtriangledown,\bigtriangleup)$, like $(+,\min)$, this seems to be very difficult, as there is no easy way to describe how each half-ancestor's $_V$ attribute changes; this is because when only some of the elements that a node $n$ covers is updated with the value $v$, how $n_V$ changes depends on what specific elements were updated, not just on how many elements were updated.

For some operator pairs, like $(+,+),(*,*)$, and $(\min,\min)$, however, this is not the case; formally, if $\exists$ a function $G$ s.t. for all sequences of numbers $[a_0,\cdots a_{k-1}]$ and for all $J\subseteq\{0,1,\cdots k-1\}$, $\bigtriangleup_{0\leq i<k} (a_i\bigtriangledown v \texttt{ if } i\in J \texttt{ else } a_i)=G(\bigtriangleup_{0\leq i<k} a_i,v,|J|,k)$, and $G$ can be calculated in constant time if its inputs are already known, then it is indeed possible to process both updates and queries in $O(polylog(N,M))$ time.

While updating the region of $n$ with the value $v$, if we pay attention to the half-ancestors $m$ s.t. $m_X\subseteq n_X,m_Y\supset n_Y$, we notice that $m_V$ becomes $G(m_V,v,|n\cap m|,|m|)=G(m_V,v,|m_X|*|n_Y|,|m|)$. A similar thing occurs for the other kind of half-ancestors. This hints at the possibility of creating new kinds of lazy values to deal specifically with half-ancestors.

Instead of using this technique, however, we exploit a property that arises with the existence of $G$ to create a more elegant algorithm: using sequences $[a,b]$ and $[a\bigtriangleup b,\emptyset_Q]$, we note that $(a\bigtriangledown v)\bigtriangleup b=G(a\bigtriangleup b,v,1,2)=G((a\bigtriangleup b)\bigtriangleup \emptyset_Q,v,1,2)=((a\bigtriangleup b)\bigtriangledown v)\bigtriangleup \emptyset_Q=(a\bigtriangleup b)\bigtriangledown v$.

This is in fact our definition of ``special": where the operators $(\bigtriangledown,\bigtriangleup)$ satisfy the requirements mentioned in Section 1.1, as well as the property

\[\bm{(a\bigtriangledown v)\bigtriangleup b=(a\bigtriangleup b)\bigtriangledown v}.\]

From this, it can be shown that $G(\bigtriangleup_{0\leq i<k} a_i,v,j,k)=(\bigtriangleup_{0\leq i<k} a_i)\bigtriangledown(v\otimes j)$, where $v\otimes j$ is an abbreviation of $(v\bigtriangledown v\bigtriangledown \cdots \bigtriangledown v)$, with the $v$ repeated $j$ times. (Note that $v\otimes j$ can be calculated in $O(\log j)$ time with binary exponentiation, or in $O(1)$ time in some cases.)

\subsection{Algorithm}
Consider a $N$-length 1D segment tree $T$, where each 1D node $n$ contains a $M$-length array $n_A$. Initialize each element $n_A[y]$ to $\bigtriangleup_{x\in n} A[x][y]$, where $A$ is the raw matrix; like before, this can be done in $O(NM)$ time using a bottom-up strategy.

Now consider doing $U(B,v)$. For each node $n\cap\not\subseteq B_X$, due to the properties that $\bigtriangledown$ and $\bigtriangleup$ have with each other, we can update $n_A$ by simply changing $n_A[y]$ to $n_A[y]\bigtriangledown (v\otimes |n\cap B_X|)$ for all $y\in B_Y$, instead of having to read $(n_l)_A$ or $(n_r)_A$. To update all nodes $\subseteq B_X$, instead of changing $n_A[y]$ to $n_A[y]\bigtriangledown (v\otimes |n|)$ for all $y\in B_Y$ and $n\subseteq B_X$, we introduce a lazy array $n_{AZ}$ for every node, where each lazy element $n_{AZ}[y]$ represents the command ``$\forall m\subseteq n$, change $m_A[y]$ to $m_A[y]\bigtriangledown (n_{AZ}[y]\otimes |m|)$". This means we can deal with all nodes $\subseteq B_X$ by changing $n_{AZ}[y]$ to $n_{AZ}[y]\bigtriangledown v$ for all $y\in [y_0,y_1]$ for all $n\in T.S(B_X)$.

Now consider doing $Q(B)$. For each node $n\in T.S(B_X)$, we must $\bigtriangleup$ together the ``true values" of $n_A[y]$ for all $y\in B_Y$. It can be shown that the true value of $n_A[y]$ is $n_A[y]\bigtriangledown (\bigtriangledown_{m\supseteq n} (m_{AZ}[y]\otimes |n|))$, so we want to calculate $\bigtriangleup_{y\in B_Y} (n_A[y]\bigtriangledown (\bigtriangledown_{m\supseteq n} (m_{AZ}[y]\otimes |n|)))$. Due to the properties of $\bigtriangledown,\bigtriangleup,$ and $\otimes$, this can be simplified to $(\bigtriangleup_{y\in B_Y} n_A[y])\bigtriangledown (\bigtriangledown_{y\in B_Y}\bigtriangledown_{m\supseteq n} (m_{AZ}[y]\otimes |n|)))=(\bigtriangleup_{y\in B_Y} n_A[y])\bigtriangledown (\bigtriangledown_{m\supseteq n}(\bigtriangledown_{y\in B_Y} m_{AZ}[y])\otimes |n|)$.

Because we are performing range $\bigtriangledown$ updates and range $\bigtriangleup$ queries on each $n_A$, and range $\bigtriangledown$ updates and range $\bigtriangledown$ queries on each $n_{AZ}$, we can convert these arrays into 1D $(\bigtriangledown,\bigtriangleup)$ segment trees and 1D $(\bigtriangledown,\bigtriangledown)$ segment trees, respectively, and change the sub-procedures of our update and query algorithms as follows:

\begin{equation}
\begin{aligned}
\big\langle n_A[y]\leftarrow n_A[y]\bigtriangledown (v\otimes |n\cap B_X|), \forall y\in B_Y\big\rangle &\Rightarrow \big\langle n_A.U(B_Y,v\otimes |n\cap B_X|)\big\rangle \\
\big\langle n_{AZ}[y]\leftarrow n_{AZ}[y]\bigtriangledown v, \forall y\in B_Y\big\rangle &\Rightarrow \big\langle n_{AZ}.U(B_Y,v)\big\rangle \\
\big\langle(\bigtriangleup_{y\in B_Y} n_A[y])\bigtriangledown (\bigtriangledown_{m\supseteq n}(\bigtriangledown_{y\in B_Y} m_{AZ}[y])\otimes |n|)\big\rangle &\Rightarrow \big\langle n_A.Q(B_Y)\bigtriangledown (\bigtriangledown_{m\supseteq n} m_{AZ}.Q(B_Y)\otimes |n|)\big\rangle
\end{aligned}
\end{equation}

We can borrow the recursive 1D segment tree update and query algorithms to make 2D updates and queries run in $O(\log N\log M)$ time, which is equal to $O(\log^2 N)$ time assuming $N\geq M$:

\begin{algorithm}[H]
\caption{Update/query a submatrix, given that $(a\bigtriangledown v)\bigtriangleup b=(a\bigtriangleup b)\bigtriangledown v$ for all numbers $a,b,v$}
\begin{algorithmic}
\Function{U}{$B,v$}
    \State UN$($root node$,B,v)$
\EndFunction
\Function{UN}{$n,B,v$}
    \If {$n\subseteq B_X$}
        \State $n_{AZ}.U(B_Y,v)$
    \ElsIf {$n$ intersects with $B_X$}
        \State UN$(n_l,B,v)$
        \State UN$(n_r,B,v)$
        \State $n_A.U(B_Y,v\otimes |n\cap B_X|)$
    \EndIf
\EndFunction
\Function{Q}{$B$}
    \State \Return QN$($root node$,B)$
\EndFunction
\Function{QN}{$n,B$}
    \If {$n\subseteq B_X$}
        \State \Return $n_A.Q(B_Y)\bigtriangledown (n_{AZ}.Q(B_Y)\otimes |n|)$
    \ElsIf {$n$ intersects with $B_X$}
        \State \Return $($QN$(n_l,B)\bigtriangleup$QN$(n_r,B))\bigtriangledown(n_{AZ}.Q(B_Y)\otimes |n\cap B_X|)$
    \Else
        \State \Return $\emptyset_Q$
    \EndIf
\EndFunction
\end{algorithmic}
\end{algorithm}

We note in Appendix \ref{special2dsegtree} that the proof of correctness of these algorithms is almost the same as that for the 1D segment tree.

Currently, we do not know if $\bigtriangledown$ and $\bigtriangleup$ can be different operators, while still satisfying the property $(a\bigtriangledown v)\bigtriangleup b=(a\bigtriangleup b)\bigtriangledown v$.

We have a Python implementation of this 2D segment tree algorithm that supports any special operator pairs at:

\centerline{\url{https://github.com/jasonLLyang/2D-Segment-Tree}}

\subsection{Generalization to multiple dimensions}
We can generalize our algorithms in the previous subsection to higher dimensions.

Suppose that $A$ is a $N\times N\times\cdots\times N$ $d$-dimensional hypermatrix for $d>1$. We use a similar technique to what we did for the 2D segment tree to create a $d$-D segment tree that performs updates and queries in $O(\log^d N)$ time.

Let $T$ be a 1D segment tree of nodes. For each node $n\in T$, let $n_A$ be a $(d-1)$-D hypermatrix where for any $(d-1)$-dimensional coordinate $c$, $n_A(c)=\bigtriangleup_{x\in n} A[x](c)$ at initialization. In addition, let $n_{AZ}$ be a $(d-1)$-D lazy hypermatrix where $n_{AZ}(c)$ represents the command ``for all $m\subseteq n$, change $m_A(c)$ to $m_A(c)\bigtriangledown (n_{AZ}(c)\otimes |m|)$".

Suppose we do $U(B,v)$. Let $X,C$ be the 1D and $(d-1)$-D submatrices that come from $B$ when the first dimension is split from the others, i.e. if $B=[l_0,r_0]\cdots[l_{d-1},r_{d-1}]$, then $X=[l_0,r_0]$ and $C=[l_1,r_1]\cdots[l_{d-1},r_{d-1}]$. For each node $n\cap\not\subseteq X$, we change $n_A(c)$ to $n_A(c)\bigtriangledown(v\otimes|n\cap X|)$ for all $c\in C$. To take care of all nodes $\subseteq X$, change $n_{AZ}(c)$ to $n_{AZ}(c)\bigtriangledown v$ for all $c\in C$ for all $n\in T.S(X)$.

Now suppose we do $Q(B)$. Letting $X,C$ be the submatrices defined like before, the true value of $n_A(c)$ is $n_A(c)\bigtriangledown(\bigtriangledown_{m\supseteq n}(m_{AZ}(c)\otimes|n|))$, so the true value of the $\bigtriangleup$ of all elements in $n*C$ is $\bigtriangleup_{c\in C}(n_A(c)\bigtriangledown(\bigtriangledown_{m\supseteq n}(m_{AZ}(c)\otimes|n|)))$, which can be simplified to $(\bigtriangleup_{c\in C}n_A(c))\bigtriangledown(\bigtriangledown_{c\in C}\bigtriangledown_{m\supseteq n}m_{AZ}(c))=(\bigtriangleup_{c\in C}n_A(c))\bigtriangledown(\bigtriangledown_{m\supseteq n}(\bigtriangledown_{c\in C}m_{AZ}(c))\otimes|n|)$.

Like before, we can convert each $n_A$ to a $(d-1)$-D $(\bigtriangledown,\bigtriangleup)$ segment tree and each $n_{AZ}$ to a $(d-1)$-D $(\bigtriangledown,\bigtriangledown)$ segment tree, and change each of the following expressions:

\begin{equation}
\begin{aligned}
\big\langle n_A(c)\leftarrow n_A(c)\bigtriangledown (v\otimes |n\cap X|), \forall c\in C\big\rangle &\Rightarrow \big\langle n_A.U(C,v\otimes |n\cap X|)\big\rangle \\
\big\langle n_{AZ}(c)\leftarrow n_{AZ}(c)\bigtriangledown v, \forall c\in C\big\rangle &\Rightarrow \big\langle n_{AZ}.U(C,v)\big\rangle \\
\big\langle(\bigtriangleup_{c\in C} n_A(c))\bigtriangledown (\bigtriangledown_{m\supseteq n}(\bigtriangledown_{c\in C} m_{AZ}(c))\otimes |n|)\big\rangle &\Rightarrow \big\langle n_A.Q(C)\bigtriangledown (\bigtriangledown_{m\supseteq n} m_{AZ}.Q(C)\otimes |n|)\big\rangle
\end{aligned}
\end{equation}

As with before, we borrow the recursive update/query algorithms of the 1D segment tree so that when updating or querying a $d$-dimensional submatrix, we only meet at most $O(\log N)$ nodes. Therefore, if $(d-1)$-D updates and queries can be done in $O(\log^{d-1} N)$ time, then $d$-D updates and queries can be done in $O(\log^d N)$ time; since this is true for $d=2$, it follows by induction that this is true for all $d$.

\subsection{Comparison to previous works}
Mishra \cite{fenw} devised a way to solve the $d$-dimensional $(+,+)$ update-query problem over a $N\times N\times\cdots\times N$ hypermatrix, with updates and queries both running in $O(4^d\log^d N)$ time; this solution takes advantage of algebraic properties of addition in order to use Fenwick trees, which are a simplified form of segment trees \cite{origfenw}. A $d$-D Fenwick tree only supports queries of the form $Q([0,r_0][0,r_1]\cdots[0,r_{d-1}])$, so querying an arbitrary submatrix involves getting multiple queries and subtracting them from each other (e.g. for $d=1$ this means that querying $[l,r]$ requires calculating $Q([0,r])-Q([0,l-1])$); this means that in general, $\bigtriangleup$ must be ``invertible", meaning that for all numbers $a$, $\exists z$ s.t. $a\bigtriangleup z=\emptyset_Q$, so this solution does not work for operator pairs like $(\min,\min)$. In contrast, our algorithm is capable of using non-invertible query operations.

Ibtehaz, Kaykobad, and Rahman \cite{mult} developed a new system of lazy values by using the techniques of ``scaling" and ``diluting", which they added on top of the traditional 2D segment tree to support submatrix updates and queries in $O(\log^d N)$ time
support more operator pairs, like $(\wedge,\wedge)$ and $(\vee,\vee)$, where $\wedge$ and $\vee$ are the AND and OR operators, respectively, over the Boolean values; note that these operators are not invertible (in fact, they lack identity values, although $(\wedge,\wedge)$ and $(\vee,\vee)$ can be simulated with $(\min,\min)$ and $(\max,\max)$ over the numbers $\in\{0,1\}$ respectively). However, their scaling and diluting of lazy values requires the use of division for the operator pair $(+,+)$ and the calculation of $n$-th roots for the operator pair $(*,*)$; as such, their algorithm uses floating-point arithmetic, even when all matrix elements and update values are integers (which is common in practice), so the query values may occasionally contain rounding errors. Additionally, their algorithm uses two types of lazy values and two types of non-lazy values for each (2D) node, which is harder to implement. In contrast, our algorithm avoids using as many extra arithmetic operations and uses 1D segment trees without having to internally modify them, making our algorithm easier to implement and immune to rounding errors (if matrix elements and update values are restricted to integers).

Below is a table comparing our 2D segment tree and the similar algorithm of \cite{mult}, showing the number of seconds it takes to process 10,000 total actions, where each action either updates or queries a random submatrix, for various $N\times N$ matrices. We used the $(+,+)$ update-query operator pair, and ran the Python implementations of both algorithms (using the implementation from \cite{mult}) on Jupyter Notebook. The main implementation of our algorithm uses Python lambdas for the update and query operators, which allow it to work for general operator pairs, but which also slow down speed, while the implementation of \cite{mult} does not, so for consistency we used a specialized implementation of our algorithm over the $(+,+)$ operator pair. The data below suggest that our algorithm is slightly faster than that of \cite{mult}.

\begin{center}
\setlength\tabcolsep{4pt}
\begin{tabular}{|c|c|c|c|c|}
\hline
 & \multicolumn{2}{|c|}{our 2D segment tree} & \multicolumn{2}{|c|}{2D segment tree from \cite{mult}} \\
\hline
$N$ & initialization time (s) & updates/queries time (s) & initialization time (s) & updates/queries time (s) \\
\hline
10 & 0.002 & 0.628 & 0.001 & 0.888 \\
30 & 0.017 & 1.583 & 0.011 & 2.264 \\
100 & 0.195 & 3.935 & 0.243 & 5.052 \\
300 & 1.700 & 7.048 & 1.698 & 9.052 \\
1000 & 17.391 & 11.714 & 18.634 & 15.752 \\
\hline
\end{tabular}
\end{center}

\section{Conclusion and Open Questions}
In this paper, we presented the hardness of the 2D update-query problem for general $(\bigtriangledown,\bigtriangleup)$ by reducing matrix multiplication to it, giving strong evidence that either update or query times cannot be truly sublinear, or preprocessing time is superpolynomial. For special $(\bigtriangledown,\bigtriangleup)$, we showed a simplified algorithm that performs both updates and queries in $O(\log N\log M)$ time. We have a few open questions:

\begin{itemize}
    \item Does there exist a data structure with polynomial-time preprocessing that supports both updates and queries in faster than $O(N)$ time for general operator pairs $(\bigtriangledown,\bigtriangleup)$ over a $N\times N$ matrix (e.g. $O(\frac{N}{\log N})$ time)?
    \item If the operators $\bigtriangledown$ and $\bigtriangleup$ satisfy $(a\bigtriangledown v)\bigtriangleup b=(a\bigtriangleup b)\bigtriangledown v$ for all numbers $a,b,v$, must $\bigtriangledown$ and $\bigtriangleup$ be the same operator?
    \item Is it possible to perform 2D updates and queries in $O(\log N\log M)$ time for operator pairs $(\bigtriangledown,\bigtriangleup)$ where $(a\bigtriangledown v)\bigtriangleup b=(a\bigtriangleup b)\bigtriangledown v$ but $\bigtriangledown$ is noncommutative? (For the 1D case, this is possible with the segment tree.)
\end{itemize}

\section{Acknowledgments}
We would like to thank Jun Wan for his mentorship and PRIMES Computer Science for their support in making this research possible.

\newpage
\appendix
\section{Proofs of correctness}
\subsection{1D segment tree}
\label{correctsegtree}
Consider a 1D segment tree over a $N$-length array. For reference, here is the algorithm to initialize the tree from the array $A$:

\begin{algorithm}[H]
\caption{Initialize segment tree}
\begin{algorithmic}
\For{each node $n$ from largest to smallest depth}
    \If{$|n|=1$}
        \State $n_V\gets A[i]$, where $n$ covers $[i,i]$
    \Else
        \State $n_V\gets (n_l)_V\bigtriangleup (n_r)_V$
    \EndIf
    \State $n_Z\gets\emptyset_U$
\EndFor
\end{algorithmic}
\end{algorithm}

For each node $n$, let $T(n)=F(n_V,\bigtriangledown_{m\supseteq n} m_Z,|n|)$ be the ``true value" of $n$. By induction, assume that this is equal to the $\bigtriangleup$ of all elements of the array that our segment tree represents that are in $n$ (i.e. $T(n)$ is correct). Then if $n$ has children, $T(n)=T(n_l)\bigtriangleup T(n_r)$ by associativity of $\bigtriangleup$. In section 2, it is shown that this rule is equivalent to the following rule: $n_V=F((n_l)_V,(n_l)_Z,|n_l|)\bigtriangleup F((n_r)_V,(n_r)_Z,|n_r|)$ (i.e. both rules imply each other); we call this the \textbf{child-value rule}. Conversely, if $T(n)=T(n_l)\bigtriangleup T(n_r)$ and $T(n)$ is correct for all $n$ s.t. $|n|=1$, then it is easy to see that $T(n)$ is correct for absolutely all $n$.

\subsubsection{Correctness of updates}
Now let us perform $U(R,v)$, using the definition of $U()$ from Algorithm 1. Let $Z(R)$ be the set of nodes whose lazy values we change, and let $V(R)$ be the set of nodes whose non-lazy values we should change after changing $n_Z$ for all $n\in Z(R)$.

\begin{algorithm}[H]
\caption{Update a subarray}
\begin{algorithmic}
\Function{U}{$R,v$}
    \State UN$($root node$,R,v)$
\EndFunction
\Function{UN}{$n,R,v$}
    \If {$n\subseteq R$}
        \State $n_Z\leftarrow n_Z\bigtriangledown v$
    \ElsIf {$n$ intersects with $R$}
        \State UN$(n_l,R,v)$
        \State UN$(n_r,R,v)$
        \State $n_V\leftarrow F((n_l)_V,(n_l)_Z,|n_l|)\bigtriangleup F((n_r)_V,(n_r)_Z,|n_r|)$
    \EndIf
\EndFunction
\end{algorithmic}
\end{algorithm}

If $n$ is not the root node, then if $UN(n,R,v)$ was called, then $UN(p,R,v)$ had to be called earlier and $p\cap\not\subseteq R$, where $p$ is the parent of $n$; consequently, all ancestors $a$ of $n$ had to be called earlier and satisfy $a\cap\not\subseteq R$.

\begin{lemma} All nodes $\in Z(R)$ are mutually disjoint.
\begin{proof}
Suppose that two different nodes $m,n\in Z(R)$ intersect; then both $UN(m,R,v)$ and $UN(n,R,v)$ were called. Because nodes in a segment tree follow a tree structure, either $m\subset n$ or $n\subset m$; WLOG let $m\subset n$; then because $UN(m,R,v)$ was called, $n\cap\not\subseteq R$, but since $n\in Z(R)$, $n\subseteq R$: contradiction.
\end{proof}
\end{lemma}

\begin{lemma} $\cup_{n\in Z(R)}n=R$.
\begin{proof}
Let $Z(n,R)$ be the set of nodes whose lazy values are changed under the $UN(n,R,v)$ call and all recursive calls that come from it. By induction, assume that $\cup_{m\in Z(n,R)}m=n\cap R$ for all $n$ with depth $>d$. Now consider any node $n$ with depth $d$. Under the $UN(n,R,v)$ call, if $n\subseteq R$, then $Z(n,R)=\{n\}$, so $\cup_{m\in Z(n,R)}m=n=n\cap R$; otherwise, if $n\cap\not\subseteq R$, then $Z(n,R)=Z(n_l,R)\cup Z(n_r,R)$, so $\cup_{m\in Z(n,R)}m=(n_l\cap R)\cup (n_r\cap R)=n\cap R$ by the induction hypothesis; otherwise, $n\cap R=\emptyset$, so $Z(n,R)=\emptyset$ and $\cup_{m\in Z(n,R)}m=\emptyset=n\cap R$. Thus, all  nodes with depth $>d-1$ satisfy the hypothesis. Because all nodes at depth $D$, where $D$ is the depth of the entire segment tree, cover exactly one element, they serve as the base case of the induction, so $\cup_{m\in Z(n,R)}m=n\cap R$ for all $n$. Because $Z(R)=Z($root node$,R)$, $\cup_{n\in Z(R)}n=$root node$\cap R=R$.
\end{proof}
\end{lemma}

For sake of clarity, let $n_{V_0}$, $n_{Z_0}$ and $T_0(n)$ ($n_{V_1}$, $n_{Z_1}$ and $T_1(n)$) represent the non-lazy, lazy and true values, respectively, of the node $n$ before (after) calling $U(R,v)$. Therefore, $n_{Z_1}=n_{Z_0}\bigtriangledown v$ if $n\in Z(R)$, $n_{Z_0}$ otherwise, and $n_{V_1}=n_{V_0}$ if $n\not\in W(R)$.

\begin{lemma} $T_1(n)$ is correct for any $n$ s.t. $|n|=1$.
\label{sing}
\begin{proof}
Here, assume that $|n|=1$. Then $n\not\in W(R)$, so $n_{V_1}=n_{V_0}$.

If $n\subseteq R$, $\exists$ one and exactly one node $n^*\in Z(R)$ s.t. $n^*$ intersects with $n$. Because $|n|=1$, $n^*\supseteq n$ and $n\not\in W(R)$, so $T_1(n)=F(n_{V_1},\bigtriangledown_{m\supseteq n} m_{Z_1},|n|)=F(n_{V_0},\bigtriangledown_{m\supseteq n} (m_{Z_0}\bigtriangledown v$ if $m=n^*$, else  $m_{Z_0}),|n|)=F(n_{V_0},(\bigtriangledown_{m\supseteq n} m_{Z_0})\bigtriangledown v,|n|)=F(F(n_{V_0},(\bigtriangledown_{m\supseteq n} m_{Z_0}),|n|),v,|n|)=F(T_0(n),v,|n|)=F(T_0(n),v,1)=T_0(n)\bigtriangledown v$, which is exactly what we want.

If $n\not\subseteq R$, then no node $\in Z(R)$ intersects with $n$, so $T_1(n)=F(n_{V_1},\bigtriangledown_{m\supseteq n} m_{Z_1},|n|)\\=F(n_{V_0},\bigtriangledown_{m\supseteq n} m_{Z_0},|n|)=T_0(n)$, which is also what we want.
\end{proof}
\end{lemma}

Whenever we change $n_V$, it is simply to restore the rule $n_V=F((n_l)_V,(n_l)_Z,|n_l|)\bigtriangleup F((n_r)_V,(n_r)_Z,|n_r|)$; therefore, we only have to do so if at least one of $(n_l)_V,(n_l)_Z,(n_r)_V$, or $(n_r)_Z$ has changed or needs to change. This means that $n\in V(R)$ iff $n_l\in V(R) \vee n_r\in V(R) \vee n_l\in Z(R) \vee n_r\in Z(R)$ (all nodes $\in V(R)$ must have children). 

\begin{lemma} $\forall n\in V(R), \ \exists m\in Z(R)$ s.t. $m\subseteq n$.
\label{VancZ}
\begin{proof}
Assume that for a number $d$, this lemma is satisfied for all $n$ with depth $>d$. Then for any node $m\in V(R)$ with depth $d$, at least one of its children must be in $V(R)$ or $Z(R)$. Let $c$ be one of those children. If $c\in Z(R)$, then the induction hypothesis is clearly satisfied; otherwise, $c\in V(R)$, and because $c$ is at a depth of $d+1$, the induction hypothesis applies, so $\exists c^*\in Z(R)$ s.t. $c^*\subseteq c \Rightarrow c^*\subseteq m$, so the hypothesis also applies to $m$; therefore, the hypothesis is true for all nodes with depth $>d-1$. Because there are no nodes at depth $D$, where $D$ is the depth of the entire tree, that acts as the base case for the induction hypothesis, so the lemma is satisfied for all node depths.
\end{proof}
\end{lemma}

This lemma means that every node whose $_V$ attribute must be updated is an ancestor of at least one of the nodes $\in Z(R)$.

\begin{lemma} $U(R,v)$ correctly updates the $_V$ attribute of every node $\in V(R)$.
\label{goodupdate}
\begin{proof}
Consider any node $n\in Z(R)$. Then for all ancestors $p$ of $n$, $UN(p,R,v)$ must have been called during $U(R,v)$, and $p\cap\not\subseteq R$, so $p_V$ is modified to follow the child-value rule. Therefore, $U(R,v)$ modifies the $_V$ attribute of every ancestor of every node $\in Z(R)$, so by Lemma \ref{VancZ} it updates the $_V$ attribute of every node $\in V(R)$. It can be proven by induction that the $_V$ and $_Z$ attributes of all nodes $\in m$ for a node $m$ will be correct after the $UN(m,R,v)$ call and all recursive calls made from that, so whenever $UN(n,R,v)$ is called and $n\cap\not\subseteq R$, $n_l$ and $n_r$ will have their $_V$ and $_Z$ attributes correct, so the resulting $n_V$ will be correct.
\end{proof}
\end{lemma}

By lemmas \ref{goodupdate} and \ref{sing}, if all $T_0(n)$ are correct, then $U(R,v)$ ensures that all $T_1(n)$ are correct. Because all $T(n)$ are correct at initialization, this leads to the following theorem:

\begin{theorem}
\label{goodT}
All $T(n)$ will always be correct after every update.
\end{theorem}

\subsubsection{Correctness of queries}
It is possible to query an arbitrary subarray $R$ by returning $\bigtriangleup_{n\in S(R)} T(n)$, which takes $O(\log^2 N)$ time to calculate, but the following recursive algorithm is faster:

\begin{algorithm}[H]
\caption{Query a subarray}
\begin{algorithmic}
\Function{Q}{$R$}
    \State \Return QN$($root node$,R)$
\EndFunction
\Function{QN}{$n,R$}
    \If {$n\subseteq R$}
        \State \Return $F(n_V,n_Z,|n|)$
    \ElsIf {$n$ intersects with $R$}
        \State \Return $F($QN$(n_l,R)\bigtriangleup$QN$(n_r,R),n_Z,|n\cap R|)$
    \Else
        \State \Return $\emptyset_Q$
    \EndIf
\EndFunction
\end{algorithmic}
\end{algorithm}

For a region $R^*$, abbreviate ``the $\bigtriangleup$ of all elements in $R^*$" as $T_\bigtriangleup(R^*)$: this acts as the ground truth.

\begin{lemma}
\label{1dquery}
$F(QN(n,R),\bigtriangledown_{m\supset n} m_Z,|n\cap R|)=T_\bigtriangleup(n\cap R)$ for any subarray $R$ and any node $n$.
\begin{proof}
By induction, assume that the lemma is true for all $n$ at depth $>d$. Now let $n$ be any node at depth $d$.

If $n\subseteq R$, then $QN(n,R)=F(n_V,n_Z,|n|)$, so $F(QN(n,R),\bigtriangledown_{m\supset n} m_Z,|n\cap R|)\\=F(F(n_V,n_Z,|n|),\bigtriangledown_{m\supset n} m_Z,|n|)=F(n_V,\bigtriangledown_{m\supseteq n} m_Z,|n|)=T(n)$, which by Theorem \ref{goodT} $=T_\bigtriangleup(n)=T_\bigtriangleup(n\cap R)$.

Otherwise, if $n\cap\not\subseteq R$, then $QN(n,R)=F(QN(n_l,R)\bigtriangleup QN(n_r,R),n_Z,|n\cap R|)$. Let $Q_L=QN(n_l,R)$ and $Q_R=QN(n_r,R)$. By the induction hypothesis, $F(Q_L,\bigtriangledown_{m\supset n_l} m_Z,|n_l\cap R|)=T_\bigtriangleup(n_l\cap R)$ and $F(Q_R,\bigtriangledown_{m\supset n_r} m_Z,|n_r\cap R|)=T_\bigtriangleup(n_r\cap R)$, so

\begin{equation*}
\begin{aligned}
F(QN(n,R),\bigtriangledown_{m\supset n} m_Z,|n\cap R|)&=F(F(Q_L\bigtriangleup Q_R,n_Z,|n\cap R|),\bigtriangledown_{m\supset n} m_Z,|n\cap R|) \\
&=F(Q_L\bigtriangleup Q_R,\bigtriangledown_{m\supseteq n} m_Z,|n\cap R|) \\
&=F(Q_L\bigtriangleup Q_R,\bigtriangledown_{m\supseteq n} m_Z,|n_l\cap R|+|n_r\cap R|) \\
&=F(Q_L,\bigtriangledown_{m\supseteq n} m_Z,|n_l\cap R|)\bigtriangleup F(Q_R,\bigtriangledown_{m\supseteq n} m_Z,|n_r\cap R|) \\
&=F(Q_L,\bigtriangledown_{m\supset n_l} m_Z,|n_l\cap R|)\bigtriangleup F(Q_R,\bigtriangledown_{m\supset n_r} m_Z,|n_r\cap R|) \\
&=T_\bigtriangleup(n_l\cap R)\bigtriangleup T_\bigtriangleup(n_r\cap R) \\
&=T_\bigtriangleup(n\cap R) \\.
\end{aligned}
\end{equation*}

Otherwise, $n\cap R=\emptyset$, so $QN(n,R)=\emptyset_Q=T_\bigtriangleup(\emptyset)=T_\bigtriangleup(n\cap R)$.

In all cases, the lemma holds for all nodes with depth $>d-1$. Since $d=D-1$, where $D$ is the depth of the entire tree, acts as a base case, the lemma holds for all nodes.
\end{proof}
\end{lemma}

From this lemma, $Q(R)=QN($root node$,R)=T_\bigtriangleup($root node$\cap R)=T_\bigtriangleup(R)$, so $Q(R)$ returns the correct value for any region $R$.

\subsection{$O(\log N\log M)$-time updates and queries for special $(\bigtriangledown,\bigtriangleup)$}
\label{special2dsegtree}

A proof of correctness for the $O(\log N\log M)$-time update/query 2D segment tree for operator pairs $(\bigtriangledown,\bigtriangleup)$ satisfying the property $(a\bigtriangledown v)\bigtriangleup b=(a\bigtriangleup b)\bigtriangledown v$ is almost identical to that of the 1D segment tree; the only difference is that some expressions need to be substituted for others as follows:

\begin{equation*}
\begin{aligned}
F(a,v,c) & \Rightarrow a\bigtriangledown(v\otimes c) \\
R & \Rightarrow B_X \\
n_V & \Rightarrow n_A.Q(B_Y) \\
n_Z & \Rightarrow n_{AZ}.Q(B_Y) \\
T(n) & \Rightarrow n_A.Q(B_Y)\bigtriangledown(\bigtriangledown_{m\supseteq n} m_{AZ}.Q(B_Y)\otimes |n|) \\
\big\langle n_V\leftarrow F((n_l)_V,(n_l)_Z,|n_l|)\bigtriangleup F((n_r)_V,(n_r)_Z,|n_r|)\big\rangle & \Rightarrow n_A.U(B_Y,v\otimes |n\cap B_X|) \\
\texttt{child-value rule} &\Rightarrow T_1(n,R)=T_1(n_l,R)\bigtriangleup T_1(n_r,R) \ \forall R, \\ \texttt{where } T_1(n,R):=n_A.Q(R)\bigtriangledown(\bigtriangledown_{m\supseteq n} m_{AZ}.Q(R)\otimes |n|)
\end{aligned} 
\end{equation*}

The 2D segment tree can be initialized with the following algorithm:
\begin{algorithm}[H]
\caption{Initialize 2D segment tree}
\begin{algorithmic}
\For{each node $n$ from largest to smallest depth}
    \If{$|n|=1$}
        \State $n_{tmp}\gets A[i]$, where $A[i]$ is the $i$-th row of $A$
    \Else
        \State $n_{tmp}\gets (n_l)_{tmp}\bigtriangleup_{\texttt{eltwise}} (n_r)_{tmp}$
    \EndIf
    \State $n_A.init(n_{tmp})$
    \State $n_{AZ}.init(E)$, where all $E[i]=\emptyset_U$ and $E$ has $M$ elements
\EndFor
\For{each node $n$}
    \State delete $n_{tmp}$
\EndFor
\end{algorithmic}
\end{algorithm}

\section{2D segment tree: $O(N\log M+M\log N)$-time updates and $\\O(\log N\log M)$-time queries for general $(\bigtriangledown,\bigtriangleup)$}
\label{general2dsegtree}
By using the idea of a ``segment tree of 1D segment trees", we can achieve a data structure that performs updates in $O(N\log M+M\log N)$ time and queries in $O(\log N\log M)$ time.

Let $T$ be the 1D segment tree over an $N$-length array. For each 1D node $n\in T$, let $n_A$ be a 1D $M$-length array s.t. $n_A[y]=\bigtriangleup_{x\in n} A[x][y]$. There will be no lazy variables that cover 1D nodes $\in T$; there will only be lazy values within each 1D node after we convert each $n_A$ to a 1D segment tree.

To update the submatrix $B$ with the value $v$, first, for each $n\subseteq B_X$, we change $n_A[y]$ to $F(n_A[y],v,|n|)$ for all $y\in B_Y$ (because the submatrix $n*[y,y]$ subsets $B$); then, for each $n\cap\not\subseteq B_X$, from largest to smallest depth, we completely reconstruct $n_A$ by setting $n_A[y]$ to $(n_l)_A[y]\bigtriangleup (n_r)_A[y]$ for all $y$ (i.e. setting $n_A$ to $(n_l)_A\bigtriangleup_{\texttt{element-wise}}(n_r)_A$).

To query the submatrix $B$, we only have to return $\bigtriangleup_{n\in T.S(B_X)} \bigtriangleup_{y\in B_Y} n_A[y]$.

Thus, for each $n_A$, we do range $\bigtriangleup$ queries and range $\bigtriangledown_{|n|}$ updates, where $a\bigtriangledown_x v:=F(a,v,x)$. Due to the properties of $F()$ noted earlier, $\bigtriangledown_x$ is commutative and associative for all $x>0$, and for any sequence of numbers $[a_0,\cdots a_{k-1}]$, we can write $\bigtriangleup_{0\leq i<k} (a_i\bigtriangledown_x v)=\bigtriangleup_{0\leq i<k} F(a_i,v,x)=F(\bigtriangleup_{0\leq i<k} a_i,v,x*k)=F_x(\bigtriangleup_{0\leq i<k} a_i,v,k)$, where $F_x(a,v,k)=F(a,v,x*k)$ is the ``$F()$" of the operator pair $(\bigtriangledown_x,\bigtriangleup)$. Because of this, we can change each $n_A$ to a 1D $(\bigtriangledown_{|n|},\bigtriangleup)$ segment tree and change the following update and query subroutines as follows:

\begin{equation*}
\begin{aligned}
\big\langle n_A[y]\gets F(n_A[y],v,|n|) \ \forall y\in B_Y \big\rangle
&\Rightarrow \big\langle n_A.U(B_Y,v) \big\rangle \\
\big\langle \bigtriangleup_{y\in B_Y} n_A[y] \big\rangle
&\Rightarrow \big\langle n_A.Q(B_Y) \big\rangle
\end{aligned}
\end{equation*}

After this transformation, queries run in $O(\log N\log M)$ time. 

The immediate approach to creating the array $C$ that $n_A$ represents is to do $C[i]=n_A.Q([i,i])$ for all $0\leq i<M$ and takes $O(M\log M)$ time. This can be improved to $O(M)$ time by performing a depth-first traversal over the nodes of $n_A$, where at every node $m$ the traversal is currently on, the algorithm stores the $\bigtriangledown_{|n|}$ of the lazy values of all ancestors of $m$.

In the final update algorithm, at most $O(N)$ different $n_A$ are updated with the $.U()$ function and at most $O(\log N)$ different $n_A$ are completely reconstructed, so the update algorithm runs in $O(N\log M+M\log N)$ time.

In $d$ dimensions, this algorithm generalizes to $O(N^{d-1}\log N)$-time updates and $O(\log^d N)$-time queries over a $N^d$-sized hypermatrix.

\subsection{Proof of correctness}

The update and query algorithms are shown below. Each node $n$ holds a 1D segment tree over the operator pairs $(\bigtriangledown_{|n|},\bigtriangleup)$, where $a\bigtriangledown_{|n|}b:=F(a,b,n)$.

\begin{algorithm}[H]
\caption{Update/Query a submatrix}
\begin{algorithmic}
\Function{U}{$B,v$}
    \ForAll{nodes $n$ from largest to smallest depth}
        \If {$n\subseteq B_X$}
            \State $n_A.U(B_Y,v)$
        \ElsIf {$n$ intersects with $B_X$}
            \State $n_A.init((n_l)_A.arr()\bigtriangleup_{\texttt{element-wise}}(n_r)_A.arr())$
        \EndIf
    \EndFor
\EndFunction
\Function{Q}{$B$}
    \State \Return QN$($root node$,B)$
\EndFunction
\Function{QN}{$n,B$}
    \If {$n\subseteq B_X$}
        \State \Return $n_A.Q(B_Y)$
    \ElsIf {$n$ intersects with $B_X$}
        \State \Return QN$(n_l,B)\bigtriangleup$QN$(n_r,B)$
    \Else
        \State \Return $\emptyset_Q$
    \EndIf
\EndFunction
\end{algorithmic}
\end{algorithm}

Let $T_\bigtriangleup(B)$ represent the ground truth $\bigtriangleup$ of all elements under $B$.

Our main goal is to prove that after any sequence of updates, for any subarray $R$, $n_A.Q(R)=T_\bigtriangleup(n*R)$. Because we have already proven that a 1D segment tree is correct over the array that it represents, we only have to prove our goal for all $R=[i,i]$. 

Suppose we do the update $U(B,v)$. Let $n_{A_0}$ ($n_{A_1}$) be $n_A$ before (after) this update, and let $T_{\bigtriangleup_0}(B)$ ($T_{\bigtriangleup_1}(B)$) be $T_\bigtriangleup(B)$ before (after) this update. By induction, assume that $n_{A_0}$ is correct (i.e. for any $n,R$, $n_{A_0}.Q(R)=T_{\bigtriangleup_0}(n*R)$.

\begin{lemma} If all $n_{A_0}$ are correct, then after $U(B,v)$, all $n_{A_1}$ will be correct.
\begin{proof}
Pick any node $n$. Since the update function goes through all nodes from largest to smallest depth, assume by induction that all nodes that have been visited before $n$ have had their $_A$ attributes modified correctly (i.e. for all $m$ visited before $n$, $m_{A_1}$ is correct).

If $n\subseteq B_X$, then the update function does $n_A.U(B_Y,v)$. This means that for all subarrays $R$ s.t. $|R|=1$, $n_{A_1}.Q(R)=F(n_{A_0}.Q(R),v,|n|)$ if $R\subseteq B_Y$, else $n_{A_0}.Q(R)$. In the first case, $F(n_{A_0}.Q(R),v,|n|)=F(T_{\bigtriangleup_0}(n*R),v,|n|)$, which $=T_{\bigtriangleup_1}(n*R)$ since $n*R\subseteq B$; in the second case, $n_{A_1}.Q(R)=T_{\bigtriangleup_0}(n*R)$, which also $=T_{\bigtriangleup_1}(n*R)$ since $n*R\cap B=\emptyset$. Therefore, $n_{A_1}.Q(R)=T_{\bigtriangleup_1}(n*R)$ for all general $R$.

Otherwise, if $n\cap B_X=\emptyset$, then $n_{A_1}=n_{A_0}$, so for all subarrays $R$, $n_{A_1}.Q(R)=n_{A_0}.Q(R)$, which is correct because $n*R\cap B=\emptyset$.

Otherwise, $n_{A_1}$ will be initialized with the element-wise $\bigtriangleup$ of $(n_l)_{A_1}$ and $(n_r)_{A_1}$; this means that for any subarray $R$, $n_{A_1}.Q(R)=(n_l)_{A_1}.Q(R)\bigtriangleup (n_r)_{A_1}.Q(R)$. By the induction hypothesis, this is equal to $T_{\bigtriangleup_1}(n_l*R)\bigtriangleup T_{\bigtriangleup_1}(n_r*R)=T_{\bigtriangleup_1}(n*R)$.

Thus, if all nodes visited before $n$ have their $_A$ attributes modified correctly, then $n_A$ will also be modified correctly. Since the first node visited in $U(B,v)$ acts as a base case, all $n_{A_1}$ are correct.
\end{proof}
\end{lemma}

Because all $n_A$ are correct at initialization, this means that after any update, all $n_A$ will still be correct.

To prove that $Q(B)=T_\bigtriangleup(B)$ for any $B$, we can prove that $QN(n,B)=T_\bigtriangleup(|n\cap B_X|*B_Y)$, using an almost identical technique to that of Lemma \ref{1dquery}, except that the induction hypothesis is swapped out with this statement.

\subsubsection{The $arr()$ function}
Given a segment tree $t$ that represents a $M$-length array, we can retrieve the array it represents using a depth-first search strategy:

\begin{algorithm}[H]
\caption{Retrieve array of 1D segment tree}
\begin{algorithmic}
\Function{$t.arr$()}{}
    \State $A\gets M$-length array
    \State $dfs(A,$root node$,\emptyset_U)$
    \State \Return $A$
\EndFunction
\Function{$dfs$}{$A,n,z$}
    \State $z_1\gets z\bigtriangledown n_Z$
    \If{$|n|=1$}
        \State $A[i]\gets n_V\bigtriangledown z_1$, where $n$ covers $[i,i]$
    \Else
        \State $dfs(A,n_l,z_1)$
        \State $dfs(A,n_r,z_1)$
    \EndIf
\EndFunction
\end{algorithmic}
\end{algorithm}

\begin{lemma} When $dfs(A,n,z)$ is called, $z=\bigtriangledown_{m\supset n}m_Z$.
\begin{proof}
Assume that this lemma is true for all $n$ at a depth $d$. Then, within a $dfs(A,n,z)$ call, if $|n|>1$, $dfs(A,n_l,z_1)$ and $dfs(A,n_r,z_1)$ where $z_1=z\bigtriangledown n_Z=\bigtriangledown_{m\supseteq n}m_Z=\bigtriangledown_{m\supset n_l}m_Z=\bigtriangledown_{m\supset n_r}m_Z$, so the lemma is true for all nodes at depth $d+1$. Since $z=\emptyset_U$ when $n$ is the root node, the lemma is true for all nodes at depth 0, so it is true for all nodes.
\end{proof}
\end{lemma}

As a consequence, if $|n|=1$, then $A[i]=n_V\bigtriangledown z_1=n_V\bigtriangledown(z\bigtriangledown n_Z)=n_V\bigtriangledown(\bigtriangledown_{m\supseteq n}m_Z)=F(n_V,\bigtriangledown_{m\supseteq n}m_Z,\allowbreak 1)=F(n_V,\bigtriangledown_{m\supseteq n}m_Z,|n|)=t.T(n)=t.Q([i,i])$, where $n$ covers $[i,i]$, and this is exactly what is desired.

Because the $arr()$ function will visit each node of $t$ exactly once, and because $t$ must have $O(M)$ nodes, the $arr()$ function runs in $O(M)$ time; this is an improvement over calling $t.Q([i,i])$ for each $i$, which takes $O(M\log M)$ time.

\section{Tightness of Lower Bound}
\label{tightness}
The quadtree is one generalization of the 1D segment tree that performs updates and queries in $O(N)$ time over a $N\times N$ matrix. It is defined as follows:

\begin{enumerate}
    \item There is a root node that covers $[0,N-1][0,N-1]$ (the entire matrix)
    \item Every node $n$ that covers $[x_0,x_1][y_0,y_1]$ s.t. $|n|>1$ has 4 children (excluding edge cases), each covering $[x_0,x_m][y_0,y_m]$, $[x_m+1,x_1][y_0,y_m]$, $[x_0,x_m][y_m+1,y_1]$, and $[x_m+1,x_1][y_m+1,y_1]$, where $x_m=\lfloor(x_0+x_1)/2\rfloor$ and $y_m=\lfloor(y_0+y_1)/2\rfloor$
    \item Every node $n$ has a value $n_V$ that is equal to $\bigtriangleup_{(x,y)\in n}A[x][y]$ at initialization
    \item Every node $n$ has a lazy value $n_Z$ that is equal to $\emptyset_U$ at initialization and represents the command ``$\forall m\subseteq n$, change $m_V$ to $F(m_V,n_Z,|m|)$
\end{enumerate}

It can be shown that there are $O(N^2)$ nodes and initialization can be done in $O(N^2)$ time with a bottom-up strategy.

Because a quadtree is is very similar to a 1D segment tree (in fact, it is equivalent to a 4-ary 1D tree over an array of all the elements in $A$ traversed in a recursively-defined path), there exist recursive update and query algorithms for it that are similar to those of the 1D segment tree, where the running time of each update or query is asymptotically equivalent to the total number of nodes visited during the update/query (and where every node will be visited at most once). Both algorithms can be roughly described like the following:

\begin{algorithm}[H]
\caption{Outline of update/query algorithms or quadtree}
\begin{algorithmic}
\Function{$\texttt{UPDATE/QUERY}$}{$B$}
    \State $\texttt{RECURSIVE}($root node$,B)$
\EndFunction
\Function{$\texttt{RECURSIVE}$}{$n,B$}
    \If{$n\subseteq B$}
        \State do some constant-time action
    \ElsIf{$n$ intersects with $B$}
        \For{$n_c$ as each child of $n$}
            \State call $\texttt{RECURSIVE}(n_c,B)$ and receive return value if any
        \EndFor
        \State do some constant-time action
    \Else
        \State do some constant-time action
    \EndIf
\EndFunction
\end{algorithmic}
\end{algorithm}

As can be seen, whenever updating or querying a submatrix $B$, the algorithm visits all nodes $\in S(B)$, where $S(B)$ is the smallest set of nodes s.t. $\cup_{n\in S(B)}n=B$ and all nodes in $S(B)$ are mutually disjoint, along with all ancestors of all nodes $\in S(B)$, and a few nodes that are disjoint from $B$.

We can ignore this last category of nodes by doing the following: first, let $I(B)=\cup_{n\in S(B)}\{m|m\supseteq n\}$; then, every node visited while updating $B$ that is disjoint to $B$ must have their parent in $I(B)$ (clear from the algorithm), and because each node has at most 4 children, there will be at most $4I(B)$ of them; thus, the number of all nodes that will be visited while updating or querying $B$ is at most $5I(B)$. Since we do not care about constant factors, we only have to pay attention to the nodes in $I(B)$.

\begin{theorem} $|I(B)|=O(N)$ for all submatrices $B$ in a $N\times N$ matrix.
\begin{proof}
We first prove the theorem for all $N=2^k$.

When paying attention to how every possible non-empty submatrix can interact with the edges of the $N\times N$ matrix, there are six types:
\begin{itemize}
    \item 0: not touching any edge
    \item 1: touching one edge
    \item 2a: touching two adjacent edges
    \item 2b: touching two opposite edges
    \item 3: touching 3 edges
    \item 4: touching all 4 edges (only occurs if the submatrix covers the entire matrix)
\end{itemize}

Define $T_i(k)$ for $i\in\{$0,1,2a,2b,3,4$\}$ to be the maximum possible number of nodes visited while updating/querying a type-$i$ submatrix in a $2^k\times 2^k$ matrix.

For general $k$, we note that a quadtree over a $2^k\times 2^k$ matrix can be recursively defined as its root node adjoined with 4 inner quadtrees, each over a $2^{k-1}\times 2^{k-1}$ matrix. Thus, we can define $T_i(k)$ recursively, since when we update/query any submatrix that does not cover the entire matrix, after visiting the root node (which we must count), we recursively visit the 4 inner quadtrees, splitting the submatrix in the process. By considering all the sets of submatrices in which each type of submatrix can be split into, and using the worst-case ones, we get the following recurrence equations:

\begin{equation*}
\begin{aligned}
T_4(k)&=1 \\
T_3(k)&=1+2T_3(k-1)+2T_4(k-1)=3+2T_3(k-1) \\
T_{2b}(k)&=1+\max(2T_{2b}(k-1),4T_3(k-1)) \\
T_{2a}(k)&=1+T_4(k-1)+2T_3(k-1)+T_{2a}(k-1)=2+2T_3(k-1)+T_{2a}(k-1) \\
T_1(k)&=1+\max(T_1(k-1)+T_{2b}(k-1),2T_3(k-1)+2T_{2a}(k-1)) \\
T_0(k)&=1+\max(T_0(k-1),2T_1(k-1),4T_{2a}(k-1))
\end{aligned}
\end{equation*}

To avoid having to deal with edge cases that occur with small $N$, we only consider $k\geq 2$, and note that $T_i(2)\leq 1+4+16=21$, which is the total number of nodes in a quadtree over a $4\times 4$ matrix. From this bound, we can show the following upper bounds via induction:

\begin{equation*}
\begin{aligned}
T_3(k)&\leq 2^{k+3}-3 \\
\Rightarrow T_3(k+1)&=3+2T_3(k) \\
&\leq 3+2(2^{k+3}-3)=2^{k+4}-3 \\
T_{2b}(k)&\leq 2^{k+4}-1 \\
\Rightarrow T_{2b}(k+1)&=1+\max(2T_{2b}(k),4T_3(k)) \\
&\leq 1+\max(2(2^{k+4}-1),4(2^{k+3}-3))=2^{k+5}-1 \\
T_{2a}(k)&\leq 2^{k+4} \\
\Rightarrow T_{2a}(k+1)&=2+2T_3(k)+T_{2a}(k) \\
&\leq 2+2(2^{k+3}-3)+2^{k+4}=2^{k+4}-4+2^{k+4} \\
&\leq 2^{k+5} \\
T_1(k)&\leq 2^{k+5}+1 \\
\Rightarrow T_1(k+1)&=1+\max(T_1(k)+T_{2b}(k),2T_3(k)+2T_{2a}(k)) \\
&\leq 1+\max(2^{k+5}+1+2^{k+4}-1,2(2^{k+3}-3)+2(2^{k+4})) \\
&\leq 1+\max(2^{k+5}+2^{k+4},2^{k+6})=2^{k+6}+1 \\
T_0(k)&\leq 2^{k+5}+3 \\
\Rightarrow T_0(k+1)&=1+\max(T_0(k),2T_1(k),4T_{2a}(k)) \\
&\leq 1+\max(2^{k+5}+3,2^{k+6}+2,2^{k+6})=2^{k+6}+3
\end{aligned}
\end{equation*}

Since the upper bounds are true for $k=2$, they are true for all $k$. Thus, $T_i(k)=O(2^k)$ for all $i$, so for $N=2^k\geq 4$, $|I(B)|\leq\max_i T_i(k)=O(2^k)=O(N)$ for any submatrix $B$ in a $N\times N$ matrix. For $N$ that are not powers of two or $N<4$, the $N\times N$ matrix can be extended to a $2^k\times 2^k$ matrix, where $k=\max(2,\lceil\log_2 N\rceil)$; since $2^{\max(2,\lceil\log_2 N\rceil)}=O(N)$, the theorem is true for all $N$.
\end{proof}
\end{theorem}

From this theorem, the update and query operations on a quadtree run in $O(N)$ time, showing that our main impossibility result is tight up to $o(N^\varepsilon)$ factors for any constant $\varepsilon>0$.

\section{Impossibility results for specific segment-tree-like algorithms to the 2D problem}
\label{nodesbad}
In this section, we show that several generalizations of the 1D segment tree require $\Omega(\sqrt{N})$ or $\Omega(N)$ time on either updates or queries (or both) in the worst case over a $N\times N$ matrix, for general $(\bigtriangledown,\bigtriangleup)$. All lower bounds presented here rely on the extended pigeonhole principle and do not depend on any unsolved problems, unlike the main impossibility results of this paper.

In all cases, there is only one kind of lazy value allowed for each node $n$: the lazy value $n_Z$ that covers all nodes $\subseteq n$. Therefore, these lower bounds show that an efficient data structure must use a different kind of lazy value or something more than just nodes. We suspect that this is not possible for most $(+,\min)$.

\subsection{Quadtrees and other trees of 2D nodes}
\label{quadtree}
One generalization of the 1D segment tree is the quad-tree: just as the segment tree starts with a root node covering the entire matrix and recursively splits each node into two halves, the quad-tre starts with a root node covering the entire matrix and (ignoring edge cases) splits every node into 4 quarters, by cutting halfway along the x- and y-axes.

Unfortunately, over a $N\times N$ matrix, representing an arbitrary submatrix with the nodes of a quad-tree requires $\Omega(N)$ nodes in the worst case, forcing updates and queries to take up to $\Omega(N)$ time: this bound is achieved with any submatrix that is a single row or a single column. This is because every node in the quadtree must cover a roughly square submatrix (if $N=2^k$ then every node must cover an exactly square submatrix), so the quadtree behaves poorly when trying to represent highly non-square submatrices (in fact, representing the submatrix $[0,N-2][0,N-2]$ also requires $\Omega(N)$ nodes). Over a $N\times M$ matrix, the quadtree encounters the same problem, as long as $N\sim M$.

Perhaps a different tree avoids this problem? To find out, we must define a tree. Formally, a set of nodes over any $d$-D hypermatrix $A$ is a \textit{tree}, or has a \textit{tree structure}, if it satisfies the following requirements:
\begin{enumerate}
    \item There exists a root node that covers all elements in the $A$.
    \item Every node that covers $>$1 element must have at least two children nodes that each cover at least 1 element.
    \item Every node that covers exactly 1 element must not have any children.
    \item All children of the same node are mutually disjoint.
    \item The union of all children of a node $n$ is equal to $n$.
\end{enumerate}

Note that each node does not have to cover a submatrix; it can cover a set of completely disconnected nodes.

We first prove an important lemma about any tree of nodes:

\begin{lemma}
\label{supsub}
For any two nodes $m$ and $n$ in a tree of nodes where $m\cap n\neq\emptyset$, $m\subseteq n$ or $n \subseteq m$.
\begin{proof}
If $m=n$ (i.e. $m$ and $n$ cover the exact same set of nodes), then the lemma is clear. From this point onward in the proof, assume $m\neq n$.

Label each node with a ``depth": the root node has a depth of 0, and if a node $n$ has depth $d$, then all of its children have depth $d+1$.

By induction, assume that all nodes at depth $d$ are mutually disjoint. Let $S=\{n_0,n_1,\ldots,n_{k-1}\}$ be the set of nodes at depth $d$ that have children and $C_i$ be the set of children of $n_i$. Then the set of all nodes with depth $d+1$ is $\cup_{0\leq i<k} C_i$. By definition of a tree, all nodes within any $C_i$ are mutually disjoint. Because $n_i$ and $n_j$ are disjoint for any $i\neq j$, any $a\in C_i$ is mutually disjoint to any $b\in C_j$, so all nodes at depth $d+1$ are mutually disjoint. Since there is only one node at depth 0, the induction hypothesis is true for $d=0$, so it is true for all $d\geq 0$. Therefore, if $m$ and $n$ are at the same depth, the lemma holds.

Now suppose $m$ and $n$ have different depths. WLOG, let $m$ have a smaller depth than $n$. Let the depth of $m$ be $d$. Then there exists a node $p$ at depth $d$ s.t. $p\supset n$, which is obtained by visiting the parent of $n$, the parent of that node, etc. and repeating this process until we are a depth of $d$. Because no two nodes at the same depth intersect, $m$ and $p$ must be the same node, so $m\supset n$.
\end{proof}
\end{lemma}

It turns out that for any 2D tree of nodes, there is a strong lower bound on how many of its nodes are required for representing an arbitrary submatrix in the worst case:

\begin{theorem}
\label{badquadtree}
For any 2D tree of nodes, there exists a submatrix that requires $\geq \frac{NM}{N+M}$ nodes to be represented.
\begin{proof}
For any $x,y$ s.t. $0\leq x<N,0\leq y<M$, consider the submatrices $B_0=[x,x][0,M-1]$ and $B_1=[0,N-1][y,y]$, and the sets $S_i=\{n$ s.t. $|n|>1,n\subseteq B_i,n\supseteq [x,x][y,y]\}, i\in \{0,1\}$.

If $|S_0|>0$ and $|S_1|>0$, then for any nodes $n_0\in S_0,n_1\in S_1$, $n_0\cap n_1\neq\emptyset$ but $n_0\not\subseteq n_1$ and $n_1\not\subseteq n_0$, which by Lemma \ref{supsub} is impossible. Therefore, either $|S_0|=0$, $|S_1|=0$, or both, so any node representation of $B_0$ or $B_1$ (i.e. any set of nodes s.t. their union is equal to $B_0$ or $B_1$) must include the node that only covers $[x,x][y,y]$.

Applying this analysis for all $x$ and $y$, there are $NM$ different $[x,x][y,y]$ but only $N+M$ different 1xN or 1xM submatrices. By the extended pigeonhole principle, there must exist a submatrix s.t. any possible node representation of it must contain $\geq \frac{N M}{N+M}$ 1x1 nodes.
\end{proof}
\end{theorem}

From this theorem, the worst case running time of both the $U()$ and $Q()$ operations using a tree of nodes is at least $\Omega(\frac{N M}{N+M})$; when $N=M$, this simplifies to $\Omega(\frac{N^2}{2N})=\Omega(N)$ time.

\subsection{Sets of arbitrary 2D nodes}
In this subsection we prove that even if the 2D segment tree (as a ``product" of 1D segment trees) is generalized to a set of arbitrary 2D nodes (none of which have to cover exact submatrices), there are still strong lower bounds similar to those of the generalized quadtree.

The only two rules of a set of arbitrary 2D nodes are that no two nodes cover the exact same set of elements, and that for every coordinate $(x,y)$ there exists a node that only covers that coordinate.

When we update a node $n$, we change its lazy value and then update the non-lazy values of all nodes $m\cap\not\subseteq n$ (to avoid visiting nodes multiple times, we will define the update function slightly differently). When we query, we do the same thing as we did with the 2D segment tree. The update and query algorithms over an arbitrary submatrix $B$ are as follows:

\begin{algorithm}[H]
\caption{Update/query a submatrix using a set of arbitrary 2D nodes}
\begin{algorithmic}
\Function{$S_U$}{$B$}
    \State \Return $\{$nodes $n_i\}$ s.t. $\cup n_i=B$ and all $n_i$ are mutually disjoint if $\bigtriangledown$ is not idempotent
\EndFunction
\Function{$S_Q$}{$B$}
    \State \Return $\{$nodes $n_i\}$ s.t. $\cup n_i=B$ and all $n_i$ are mutually disjoint if $\bigtriangleup$ is not idempotent
\EndFunction
\State ($S_U()$ and $S_Q()$ must return the same set of nodes every time they receive the same input submatrix $B$)
\Function{U}{$B,v$}
    \ForAll{$n \in S_U(B)$}
        \State $n_Z\gets n_Z\bigtriangledown v$
    \EndFor
    \ForAll{$m$ where $\exists n\in S_U(B)$ s.t. $n\cap m\neq\emptyset$ but $\nexists n\in S_U(B)$ s.t. $n_Z$ covers $m$}
        \State change $m_V$ (how we actually do this is not important)
    \EndFor
\EndFunction
\Function{Q}{$B$}
    \State \Return $\bigtriangleup_{n \in S_Q(B)} F(n_V,\bigtriangledown_{m \supseteq n} m_Z,|n|)$
\EndFunction
\end{algorithmic}
\end{algorithm}

Let $C(B)$ denote the set $\{m|m\cap\not\subseteq B\}$ for any submatrix $B$.

\begin{lemma} For any submatrix $B$, every node $m\in C(B)$ is visited during $U(B,v)$.
\begin{proof}
$\forall m$ s.t. $m\cap\not\subseteq B$, $\exists n\in S_U(B)$ s.t. $n$ intersects with $m\cap B$, since $\cup_{n\in S_U(B)}n=B$. Because $n\subseteq B$, $m\cap\not\subseteq n$, so $m$ will be visited by $U(B,v)$.
\end{proof}
\end{lemma}

We now present a theorem similar to Theorem \ref{badquadtree}:
\begin{theorem} The $U()$ function must visit $\geq \frac{NM}{N+M}$ nodes in the worst case.
\begin{proof}
For integers $x,y$ s.t. $0\leq x<N, 0\leq y<M$, define submatrices $B_0=[x,x][0,M-1],B_1=[0,N-1][y,y]$. Then there must exist a node $n\in S_U(B_0)$ s.t. $n\supseteq [x,x][y,y]$. Choose any such $n$. If $|n|=1$, mark it as a special contributor to $S_U(B_0)$. Otherwise, $n\in C(B_1)$, so mark it as a special contributor to $C(B_1)$. Apply this logic for all $x$ and $y$.

Since every node marked as special contributor to $S_U(B_0)$ only covers a single element, and that element is different for every pair of values $(x,y)$, no node is marked as a special contributor to the same $S_U(B_0)$ twice. Furthermore, every node marked as a special contributor to the same $C(B_1)$ is within a different x-slice of the matrix, so no node is counted as a special contributor to the same $C(B_1)$ twice. Therefore, $\sum_{B\in T} |S_U(B)|+|C(B)| \geq NM$, where $T=\{[x,x][0,M-1] \ \forall 0\leq x<N\}\cup \{[0,N-1][y,y] \ \forall 0\leq y<M\}$. Since $|T|=N+M$, by the extended pigeonhole principle, $\exists B\in T$ s.t. $|S_U(B)|+|C(B)| \geq \frac{NM}{N+M}$, which is no greater than how many nodes $U(B,v)$ visits.
\end{proof}
\end{theorem}

When $N=M$, this means that the $U()$ function runs in $\Omega(\frac{N^2}{2N})=\Omega(N)$ worst-case time. Note that this theorem does not necessarily apply to $Q()$, as the 2D segment tree we presented earlier is a counterexample.

Suppose that during $U(B,v)$, we decide not to update the $_V$ attributes of all necessary nodes. For that, we first redefine the $U()$ and $Q()$ operations:
\begin{algorithm}[H]
\caption{Update/query a submatrix}
\begin{algorithmic}
\Function{U}{$B,v$}
    \State let $S_U,S_C$ be sets of nodes where $\cup_{n\in S_U} n=B$
    \ForAll{$n \in S_U$}
        \State $n_Z\gets n_Z\bigtriangledown v$
    \EndFor
    \ForAll{$m\in S_C$}
        \State change $m_V$ (how we actually do this is not important)
    \EndFor
\EndFunction
\Function{Q}{$B$}
    \State let $S_Q$ be a set of nodes s.t. $\cup_{n\in S_Q} n=B$
    \State \Return $\bigtriangleup_{n \in S_Q} F(n_V,\bigtriangledown_{m \supseteq n} m_Z,|n|)$
\EndFunction
\end{algorithmic}
\end{algorithm}

Define $N_C(x,y)=\{n$ s.t. $|n|>1,n\subseteq [0,N-1][y,y],n\supseteq [x,x][y,y]\}$.

\begin{lemma}
\label{q1x1}
For general $(\bigtriangledown,\bigtriangleup)$ operator pairs, and for all $0\leq x<N,0\leq y<M$, if, while updating $[x,x][0,M-1]$, $S_C\not\ni n$ for all $n\in N_C(x,y)$, then when querying $[0,N-1][y,y]$ immediately afterward, $S_Q$ must contain the node that only covers $[x,x][y,y]$, and must not contain any nodes in $N_C(x,y)$.
\begin{proof}
Let $(\bigtriangledown,\bigtriangleup)=(+,\min)$. Initialize the element at row $x$, column $y$ to 0, and all other elements of the matrix to $I$, where $I>0$. This means that for all $n$, $n_V=(0$ if $n\supseteq [x,x][y,y]$ else $I)$ and $n_Z=0$.

Now do $U(B_0,v)$ for any $0<v<I$, and then do $Q(B_1)$; let $S_Q$ be the set $S_Q$ that was determined during the $Q()$ operation call. Suppose that for all $n\in N_C(x,y), n_V$ is never updated. Then for such $n, n_V=0$ instead of $v$. If $S_Q\ni n$, it will evaluate $n_V+\sum_{m\supseteq n} m_Z$; since $|n|>1$, for all $m\supseteq n, m\not\subseteq B_0$, so $m\not\in S_U$ and $m_Z$ cannot be updated during $U(B_0,v)$ (if $m_Z$ was updated, the update would cover elements outside of $B_0$), so this expression will $=0$ and thus force $Q(B_1)$ to return 0, when the correct answer is $v$. Therefore, $S_Q\not\ni n$ for all $n\in N_C(x,y)$, so $S_Q$ must contain a node that only covers $[x,x][y,y]$ (and such a node can be covered by a lazy update of a node $m\in B_0$). 
\end{proof}
\end{lemma}

\begin{theorem} If $S_U,S_C,S_Q$ are each fixed functions over submatrices (i.e. each one of them always returns the same set of nodes when given the same input submatrix), then $U()$ or $Q()$ (or both) must visit $\geq \frac{NM}{N+M}$ nodes in the worst case.
\begin{proof}
For integers $x,y$ s.t. $0\leq x<N, 0\leq y<M$, let $B_0=[x,x][0,M-1],B_1=[0,N-1][y,y]$ be submatrices. If $\exists n\in N_C(x,y)$ s.t. $n\in S_C(B_0)$, then we choose any such $n$ as a special contributor to $S_C(B_0)$. Otherwise, by Lemma \ref{q1x1}, $S_Q(B_1)\ni n^*$, where $n^*$ only covers $[x,x][y,y]$, so we mark $n^*$ as a special contributor to $S_Q(B_1)$.

Applying this logic for all $x$ and $y$ shows that \[\sum_{0\leq x<N} |S_U([x,x][0,M-1])|+\sum_{0\leq y<M} |S_Q([0,N-1][y,y])|\geq NM.\] By the extended pigeonhole principle, $\exists x$ s.t. $|S_U([x,x][0,M-1])|\geq \frac{NM}{N+M}$ or $\exists y$ s.t. $|S_Q([0,N-1][y,y])|\geq \frac{NM}{N+M}$, so at least one $U()$ or one $Q()$ operation must take $\Omega(\frac{NM}{N+M})$ time in the worst case.
\end{proof}
\end{theorem}

\begin{theorem} If $S_U,S_C,S_Q$ are not necessarily fixed functions over submatrices, then $U()$ or $Q()$ (or both) must visit $\geq \max(\min(\sqrt{M},N),\min(\sqrt{N},M))$ nodes in the worst case (assuming $N$ and $M$ are proportional to each other).
\begin{proof} Let $1\leq k\leq N$ be an integer. Then do $U([0,0][0,M-1],v_0),U([1,1][0,M-1],v_1),\cdots U([k-1,k-1][0,M-1],v_{k-1})$. Let $S^x_C$ be the set $S_C$ determined when $U([x,x][0,M-1],v_x)$ was called. If $\exists y$ s.t. for all $0\leq x<k, S^x_C\not\ni n$ for all $n\in N_C(x,y)$, then by extending Lemma \ref{q1x1}, doing $Q([0,N-1][y,y])$ means that the set $S_Q$ contains $n^*$, where $n^*$ only covers $[x,x][y,y]$, for each $x$ s.t. $0\leq x<k$, so $|S_Q|\geq k$. Otherwise, for all $y$, $\exists x$ s.t. $\exists n\in N_C(x,y)$ s.t. $S^x_C\ni n$, so $\sum_{0\leq x<k} |S^x_C|\geq M$, so $\exists x$ s.t. $|S^x_C|\geq \frac{M}{k}$. Applying this logic for all $1\leq k\leq N$ gives a $\max_{1\leq k\leq N} \min(k,\frac{M}{k})$ lower bound, which simplifies to $\max_{1\leq k\leq N} (k$ if $k<\sqrt{M}$ else $\frac{M}{k})=\max(\max_{1\leq k\leq \min(\sqrt{M},N)} k,\max_{\sqrt{M}<k\leq N} \frac{M}{k})=\min(\sqrt{M},N)$. Applying the same argument but with the x- and y- axes flipped yields a $\min(\sqrt{N},M)$ lower bound, so the final lower bound is $\max(\min(\sqrt{M},N),\min(\sqrt{N},M))$.
\end{proof}
\end{theorem}

When $M=N$, this lower bound simplifies to $\sqrt{N}$ nodes.
\end{document}